\documentclass[twocolumn,superscriptaddress,prx]{revtex4-2}
\usepackage{amsmath,amssymb}
\usepackage{graphicx}
\usepackage{color}
\usepackage{siunitx}
\usepackage{physics}
\usepackage{circuitikz}
\usepackage{float}
\usepackage{bm}

\begin{document}

\title{Dirac branch-cut modes with relativistic transport}

\author{Bofeng Zhu}
\altaffiliation{These authors contributed equally to the work.}

\affiliation{Division of Physics and Applied Physics, School of Physical and Mathematical Sciences,
Nanyang Technological University, Singapore 637371, Singapore}
\affiliation{Centre for Disruptive Photonic Technologies, Nanyang Technological University, Singapore 637371, Singapore}

\author{Chengzhi Ma}
\altaffiliation{These authors contributed equally to the work.}

\affiliation{School of Electrical Engineering, Xi'an Jiaotong University, Xi'an 710049, China}
\affiliation{Division of Physics and Applied Physics, School of Physical and Mathematical Sciences,
Nanyang Technological University, Singapore 637371, Singapore}
\affiliation{State Key Laboratory of Electrical Insulation and Power Equipment, Xi'an 710049, China}

\author{Qiang Wang}
\affiliation{School of Physics, Nanjing University, Nanjing 210093, China}

\author{Gui-Geng Liu}
\affiliation{School of Engineering, Westlake University, Hangzhou, Zhejiang 310024, China}

\author{Xiuhai Zhang}
\affiliation{Division of Physics and Applied Physics, School of Physical and Mathematical Sciences, Nanyang Technological University, Singapore 637371, Singapore}
\affiliation{School of Marine Science and Technology, Northwestern Polytechnical University, Xi'an 710072, China}

\author{Zheyu Cheng}
\affiliation{Division of Physics and Applied Physics, School of Physical and Mathematical Sciences, Nanyang Technological University, Singapore 637371, Singapore}

\author{Qi Jie Wang}
\email{qjwang@ntu.edu.sg}
\affiliation{Centre for Disruptive Photonic Technologies, Nanyang Technological University, Singapore 637371, Singapore}
\affiliation{School of Electrical and Electronic Engineering, Nanyang Technological University, Singapore 639798, Singapore}

\author{Baile Zhang}
\email{blzhang@ntu.edu.sg}
\affiliation{Division of Physics and Applied Physics, School of Physical and Mathematical Sciences, Nanyang Technological University, Singapore 637371, Singapore}
\affiliation{Centre for Disruptive Photonic Technologies, Nanyang Technological University, Singapore 637371, Singapore}

\author{Y.~D.~Chong}
\email{yidong@ntu.edu.sg}
\affiliation{Division of Physics and Applied Physics, School of Physical and Mathematical Sciences, Nanyang Technological University, Singapore 637371, Singapore}
\affiliation{Centre for Disruptive Photonic Technologies, Nanyang Technological University, Singapore 637371, Singapore}


\begin{abstract}
Emergent Dirac fields, exhibiting effective relativistic physics, are most commonly associated with bulk and surface states in materials such as graphene and topological insulators. Here we show that phase dislocations of a complex Dirac mass support a distinct family of propagating Dirac defect states that behave as massless or massive one-dimensional relativistic modes. These traveling-wave defect states, termed Dirac branch-cut (DBC) modes, are governed by a one-dimensional Dirac Hamiltonian whose effective mass is controlled by the phase difference across the branch cut. Using acoustic metamaterials, we experimentally demonstrate a range of relativistic phenomena exhibited by DBC modes, including relativistic dispersion, energy-independent confinement, Klein tunnelling, and transport along freeform (e.g., spiral) trajectories. Our results establish branch-cut defects as a distinct platform for realizing propagating Dirac defect states, extending relativistic Dirac physics from bulk and surface states to propagating modes confined to defect boundaries.
\end{abstract}

\maketitle

Dirac fields are of profound importance in many areas of physics \cite{Thaller2013}, including particle physics \cite{Gottfried1986} and condensed matter physics \cite{Wehling2014}, because they provide a natural framework for effective relativistic wave phenomena. The discovery of graphene showed that such physics is not restricted to elementary particles, but can emerge as low-energy excitations in crystalline materials \cite{Novoselov2005,Neto2009}. Since then, phenomena such as Klein tunneling \cite{Klein1929} have been explored primarily through extended bulk or surface states associated with Dirac cones \cite{Katsnelson2006,Ni2018,Jiang2020,Yu2024}. In this setting, relativistic behaviour is tied to dispersive Dirac bands and the propagation of waves through the bulk or along material surfaces.

A separate and equally influential direction concerns defect states in Dirac fields. The best-known examples are the Jackiw--Rebbi state \cite{Jackiw1976}, localized at a domain-wall defect where a real Dirac mass changes sign, and the Jackiw-Rossi state \cite{Jackiw1981}, bound to a vortex of a complex Dirac mass. These states have shaped our understanding of charge fractionalization \cite{Su1979}, topological phases of matter \cite{hasan2010colloquium, Qi2011Review}, and Majorana physics \cite{Volovik1999, Read2000,Fu2008}, and have been realized across condensed-matter and metamaterial platforms \cite{OzawaReview2019, MaReview2019,ChenReview2026,Kekule, Hou2007, Gutierrez2016, Gao2019, Noh2020, Gao2020, Menssen2020_braiding, Yang2022, Han2023}. However, while Dirac cones provide propagating relativistic states, the associated defects have been viewed mainly as mechanisms for binding topological states. Whether a lower-dimensional defect can itself host relativistic Dirac modes, and the resulting consequences, remain largely unexplored.


Here we show that a phase discontinuity in a complex Dirac mass field---a branch-cut defect---supports such states. A branch cut arises when the complex mass field is assigned to a multi-valued complex function, such as a root or logarithm, producing a phase discontinuity along a curve \cite{martin1966complex, Brown2009}. We find that this phase discontinuity binds a distinct family of propagating relativistic defect states, which we term Dirac branch-cut (DBC) modes.  Unlike Jackiw-Rebbi and Jackiw-Rossi states, DBC modes satisfy a one-dimensional (1D) Dirac equation, retaining the relativistic features of the parent Dirac field, including the same Dirac velocity.  Their effective mass is determined by the phase discontinuity across the cut, and may be zero (linear dispersion) or nonzero (hyperbolic dispersion); in either case, the DBC modes exhibit energy-independent transverse confinement length, a signature property arising from the fact that both the bulk and defect are relativistic.  We experimentally realize DBC modes using acoustic metamaterials composed of a Kekul\'e modulated triangular lattice of rigid pillars \cite{Kekule, Hou2007, Gao2019}, with the modulation amplitude controlling the mass magnitude and the modulation phase controlling the phase. The resulting DBC modes exhibit relativistic dispersion, constant transverse confinement, and Klein tunneling along the defect. We also demonstrate DBC modes in a variety of configurations, including line cavities formed by finite cuts and freeform transport along spiral trajectories. These properties establish branch cuts as a distinct platform for realizing Dirac defect states alongside domain walls and vortices, while revealing a new regime in which relativistic Dirac transport occurs along defect boundaries rather than in bulk or surface states.

To derive the existence of DBC modes, we take the Jackiw--Rossi model, which comprises a minimal four-component Dirac equation in 2D space with the Hamiltonian \cite{Jackiw1981, Gao2020}.
\begin{equation} \label{eq:hamiltonian}
  H = \tau_3 \left(-i \sigma_1 \frac{\partial}{\partial x} - i \sigma_2 \frac{\partial}{\partial y}\right)
  + m_1(\mathbf{r}) \tau_1 + m_2(\mathbf{r}) \tau_2,
\end{equation}
where the Dirac velocity is normalized to 1, the 2D position vector is $\mathbf{r} \equiv (x,y)$, and the internal Hilbert space is factorized as $\mathbb{C}^2\otimes\mathbb{C}^2$, with $\sigma_{1,2,3}$ and $\tau_{1,2,3}$ denoting Pauli matrices acting on each factor space.  There are two Dirac mass fields $m_1$ and $m_2$, which can be consolidated into a complex mass field $m(\mathbf{r}) \equiv m_1 + im_2 = |m|\,e^{i\theta}$; the model excludes other mass terms that break chiral symmetry \cite{Gao2020}.  Note that $H$ is nonetheless Hermitian.  Jackiw and Rossi considered a specific $m(\mathbf{r})$ with $\theta$ winding by $2\pi$ around the origin, and showed that it hosts a mid-gap bound state \cite{Jackiw1981}; this was later generalized to an index theorem stating that any $m(\mathbf{r})$ with a $2\pi$ phase winding implies the existence of such a bound state \cite{Weinberg1981}.

\begin{figure*}
  \centering
  \includegraphics[width=0.95\textwidth]{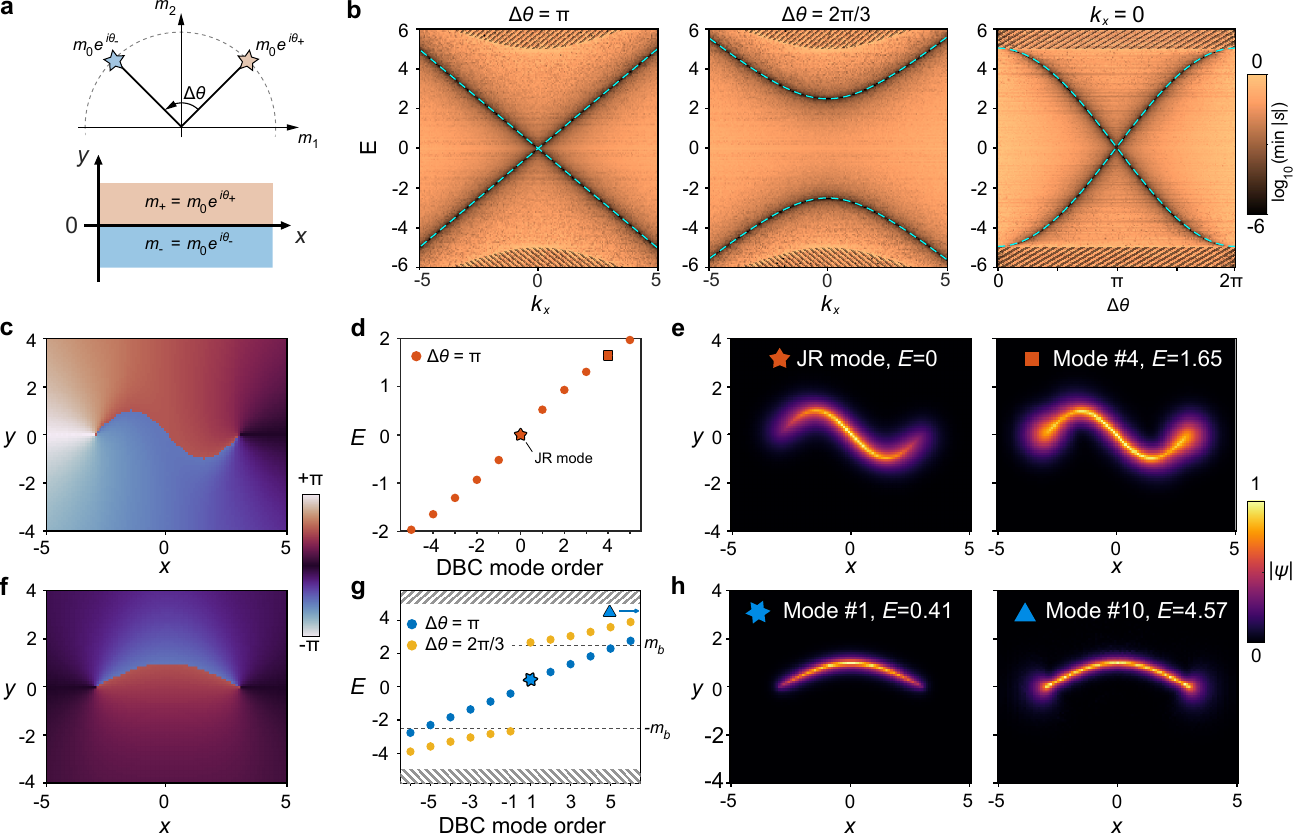}
  \caption{\textbf{Properties of Dirac branch-cut (DBC) modes.} \textbf{a}, Upper panel: complex plane for the Jackiw--Rossi mass parameter $m$, indicating two points with the same $|m|$ and phase difference $\Delta \theta$.  Lower panel: a schematic of a piecewise-constant $m(\mathbf{r})$ distribution in 2D space, constructed from those two $m$ values with phase dislocation along $y = 0$.  \textbf{b}, Phase dislocation mode (DBC mode) energies versus wavenumber $k_x$ for $\Delta\theta=\pi$ (left panel) and $\Delta\theta=2\pi/3$ (middle panel), and versus $\Delta\theta$ at $k_x = 0$ (right panel).  Cyan dashes show the results obtained analytically using Eqs.~\eqref{dispersion}--\eqref{kappa_fixed}.  Color maps show the minimum eigenvalue magnitude of the scattering matrix, computed directly from the Hamiltonian \eqref{eq:hamiltonian}; a zero value corresponds to a boundary state (see Supplementary Information).  The bulk bands are shaded in gray. \textbf{c}, Phase profile of a mass distribution made from a branch function $m = (z-3)^{1/2}(z+3)^{1/2}$, where $z = x + iy$.  \textbf{d}, Numerically-obtained DBC mode energies versus mode order with the phase of mass distribution in \textbf{c}, and the amplitude extracted from the branch function $|m|$. \textbf{e},  Spatial profiles for two of the modes in \textbf{d}: the Jackiw--Rossi state at $E = 0$ (left panel), and the order-4 mode (right panel).  \textbf{f}, Phase profile of a ``phase-only'' mass distribution $|m| = m_0$ and $\theta = \mathrm{arg}[(z-3)^{1/2}(z+3)^{-1/2}]$, with $m_0 = 5$.  \textbf{g}, DBC mode energies for the mass distribution in \textbf{f} (blue markers), and for another mass distribution with $\Delta\theta = 2\pi/3$ using the same cut (yellow markers).  The theoretically-predicted minigap for the latter is indicated by horizontal dashes.  \textbf{h} Representative spatial profiles of two of the DBC modes for $\Delta\theta=\pi$. For the DBC mode profiles for $\Delta\theta=2\pi/3$, see Supplementary Information Figs. S1c,d. All the DBC modes have very similar confinement lengths, up to high orders approaching the bulk bands. }
  \label{fig:2D_Dirac}
\end{figure*}

Now, instead of phase singularities, we study phase dislocations. Consider the piecewise-constant mass field in Fig.~\ref{fig:2D_Dirac}\textbf{a}, with $m = m_0 e^{i\theta_+}$ for $y > 0$ and $m = m_0 e^{i\theta_-}$ for $y < 0$, where $m_0 \in \mathbb{R}^+$ is a constant.  In each domain, there is a bulk gap $[-m_0, m_0]$, while at the domain wall $y = 0$ there is a phase discontinuity $\Delta \theta = \theta_+ - \theta_-$.  For any $\Delta\theta \ne 0$, there exist DBC modes along the domain wall, with the Jackiw--Rebbi-like form
\begin{align}
  &\psi(y) =
  \begin{cases}
    e^{+\kappa y}\, e^{- i \tau_3\theta_- /2} \, |\chi_-\rangle, & y<0,\\
    e^{-\kappa y}\, e^{- i \tau_3\theta_+/2} \, |\chi_+\rangle, & y>0,
  \end{cases}
  \label{eq:ansatz}
\end{align}
where $\kappa= \left(m_0^{2}+k_x^{2}-E^{2}\right)^{1/2}$ is the inverse decay depth, $k_x$ is the wavenumber along $x$, and the energy $E$ lies in the bulk gap ($|E| < m_0$), and $|\chi_\pm\rangle$ are spinors satisfying
\begin{align}
  \big[\tau_3\left(k_x\sigma_1 \pm i\kappa\sigma_2 \right)
    + m_0 \tau_1\big] |\chi_\pm\rangle = E |\chi_\pm\rangle.
  \label{eq:bulk_spinor_eq}
\end{align}
It can be shown that the continuity of $\psi(y)$ along $y = 0$ constrains the solutions to Eq.~\eqref{eq:bulk_spinor_eq} to a $\Delta\theta$-dependent subspace (see Methods).  The DBC mode eigenproblem then maps to a two-component 1D Dirac equation, with
\begin{align}
  E^2 &= k_x^2 + m_b^2, \;\;\; m_b = m_0 \left|\cos(\Delta \theta/2)\right|,
  \label{dispersion} \\
  \kappa &= m_0 \left|\,\sin(\Delta\theta/2)\,\right|.
  \label{kappa_fixed}
\end{align}
In Eq.~\eqref{dispersion}, the effective mass $m_b$ depends on $m_0$ and the phase dislocation $\Delta\theta$.  For $\Delta\theta=\pi$, which corresponds to a Jackiw--Rebbi-like sign flip of $m$ across the domain wall, we have $m_b = 0$; other values of $\Delta\theta$ give $0 < m_b < m_0$, generating a tunable family of massive 1D Dirac modes.  The addition of a phase gradient $v_\theta y$ retains this result, but with a correction to $m_b$ (see Methods).  Importantly, $\kappa$ is independent of $k_x$, so all the DBC modes have the same transverse confinement length regardless of where they lie in the bulk gap.  This is due to the 1D dispersion relation \eqref{dispersion} and the bulk dispersion relation for guided modes, $E^2 = k_x^2 - \kappa^2 + m_0^2$, from Eq.~\eqref{eq:hamiltonian}; these involve the same Dirac velocity, and combining them leads to Eq.~\eqref{kappa_fixed}.  We will return to this robust confinement property later.

In Fig.~\ref{fig:2D_Dirac}\textbf{b}, we compare the analytic result Eq.~\eqref{dispersion} to numerical results obtained directly from Eq.~\eqref{eq:hamiltonian} (see Supplementary Information, Sec.~S1).  We obtain exact agreement for the dispersion relations (left and middle panels) and their $\Delta \theta$-dependence (right panel).

From these findings, we hypothesize that the DBC modes arise chiefly from the presence of a discontinuity in $\theta(\mathbf{r})$, and will also exist for phase dislocations along arbitrary 1D curves.  A convenient way to generate phase dislocations is to use a branch-cut---a 1D curve connecting different Riemann sheets of a complex multi-valued function, like a complex root or logarithm \cite{martin1966complex}. A branch function is specified by choosing one Riemann sheet, and is differentiable for all $\mathbf{r}$ except at the cut, where there is a constant phase discontinuity given by the monodromy (i.e., the multiplicity of the Riemann sheets).  As an example, take $f(z) = (z-z_1)^ \alpha (z-z_2)^ \beta$, where $z = x + iy$, and $\alpha, \beta \in \mathbb{Q}$ with $\alpha + \beta \in \mathbb{Z}$.  The branch points are $z_1$ and $z_2$, with no branch point at infinity; we can place the cut freely so long as it does not cross itself and terminates at the branch points.  All along the cut, there is a phase discontinuity $\Delta \theta \equiv 2\pi\alpha \equiv -2\pi\beta\,(\textrm{mod}~2\pi)$. 

The cut can be continuously deformed between the branch points without changing its monodromy-induced phase discontinuity $\Delta\theta$. This phase discontinuity supports the extended DBC modes, whereas a nonzero net phase winding of the overall mass texture can additionally enforce a Jackiw-Rossi zero mode.

In Fig.~\ref{fig:2D_Dirac}\textbf{c}, we show the distribution of $\theta(\mathbf{r}) = \mathrm{arg}[m(\mathbf{r})]$ for $m(\mathbf{r}) = (z-3)^{1/2}(z+3)^{1/2}$.  The cut is chosen to be a sinusoidal curve connecting the branch points at $z = \pm 3$, and the phase discontinuity is $\Delta \theta = \pi$.  Since $|m| \rightarrow \infty$ as $|\mathbf{r}|\rightarrow \infty$, there are no free solutions, only bound states.  We solve for these numerically using the finite element method (FEM) (see Methods and Supplementary Information, Sec.~S2). The spectrum consists of discrete energies extending to $E \rightarrow \pm \infty$, as shown in Fig.~\ref{fig:2D_Dirac}\textbf{d}.  As the branch function has a phase winding of $2 \pi$ upon encircling both branch points, the index theorem for the Jackiw--Rossi model predicts a mode at $E = 0$ \cite{Jackiw1981, Weinberg1981}, consistent with our obtained spectrum. Thus, in this particular configuration, the Jackiw–Rossi zero mode coincides with the zero-energy member of the DBC spectrum.  However, all the DBC modes are strongly bound to the cut.  Two of them, the zero mode and the order-4 mode, are plotted in Fig.~\ref{fig:2D_Dirac}\textbf{e}.


\begin{figure*}
  \centering
  \includegraphics[width=0.95\textwidth]{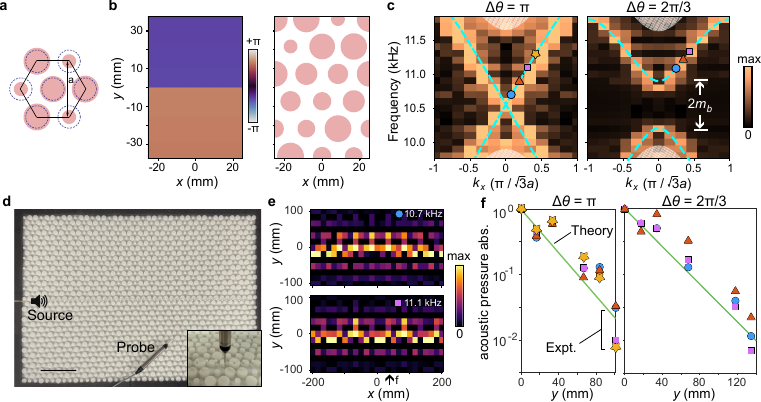}
  \caption{\textbf{Implementation of DBC modes in an acoustic structure.} \textbf{a}, Schematic of the acoustic structure, consisting of solid pillars (pink circles) surrounded by air.  The pillar positions are fixed, and their radii are modulated according to Eq.~\eqref{eq:Kekule_acoustic}; in this plot, we take $\theta = -0.35\pi$.  Dashed and solid circles indicate the radii before and after modulation, respectively. \textbf{b}, A phase map $\theta(\mathbf{r})$ with a $\pi$ phase dislocation along $y = 0$ (left panel), and the corresponding acoustic structure (right panel). The upper half is assigned $\theta = -0.35\pi$ and modulation parameters $R_0=0.2a$ and $\Delta R=0.13d$; the lower half has $\theta = 0.65\pi$, $R_0=0.19a$, and $\Delta R=0.092d$. \textbf{c}, DBC mode dispersion for $\Delta\theta = \pi$ (left panel) and $\Delta\theta = 2\pi/3$ (right panel).  Color maps show the Fourier transformed excitation spectrum obtained from experiments, while cyan dashes show the dispersion relations from FEM simulations of the acoustic structure (see Methods). The bulk bands are shaded in gray.  Markers in the left and right panels indicate four frequencies between 10.7~kHz--11.3~kHz (with 0.2~kHz spacing), and three frequencies between 11.1~kHz--11.3~kHz (with 0.1~kHz spacing) used in subsequent subplots. \textbf{d}, Experimental setup for measuring spatially-resolved excitation spectra. The location of the acoustic source is indicated by the loudspeaker symbol. \textbf{e}, Measured profiles for DBC modes at 10.7~kHz (upper panel) and 11.1~kHz (lower panel). \textbf{f}, Semilogarithmic plot of the transverse acoustic pressure profiles, measured along a line of constant $x$ with the frequencies at $\Delta\theta = \pi$ (left panel) and $\Delta\theta = 2\pi/3$ (right panel) marked in \textbf{c}.  These experimental results are normalized to the value at $y = 0$, and agree with the frequency-independent theoretical prediction derived from the Jackiw--Rossi model (solid green line).}
  \label{fig:acoustic}
\end{figure*}

We can also devise mass distributions with finite bulk gaps, which are more suited to experimental realization (see below).  An expedient way to do this is to take only the phase of a complex branch function, keeping the magnitude constant.  As an example, we take $|m| = 5$ with $\theta = \mathrm{arg}\{[(z-3)/(z+3)]^{1/2}\}$, using an arc-like cut, as shown in Fig.~\ref{fig:2D_Dirac}\textbf{f}; though this $m(\mathbf{r})$ is non-holomorphic, that does not invalidate our preceding discussion.  There is now a bulk gap $E \in [-5,5]$, with free states outside and DBC modes inside the gap (Fig.~\ref{fig:2D_Dirac}\textbf{g}, blue circles).  The net phase winding is now zero, so the DBC modes do not include a Jackiw--Rossi mode.  For comparison, we also plot the spectrum for $|m| = 5$ and $\theta = \mathrm{arg}\{[(z-3)/(z+3)]^{1/3}\}$, which has phase discontinuity $\Delta\theta = 2\pi/3$ (Fig.~\ref{fig:2D_Dirac}\textbf{g}, yellow circles).  These DBC modes exhibit a minigap closely matching the one derived from Eq.~\eqref{dispersion}, as indicated by dashes in the figure.  The DBC modes are again strongly confined to the cut, as shown in Fig.~\ref{fig:2D_Dirac}\textbf{h}.  Remarkably, even high-order DBC modes very close to the bulk band edge (e.g., mode \# 10 in Fig.~\ref{fig:2D_Dirac}\textbf{g},\textbf{h}) have the same tight confinement, consistent with the prediction from Eq.~\eqref{kappa_fixed}.

In the Supplementary Information, Sec.~S3, we investigate the mode profiles in greater detail.  All DBC modes have profiles matching those in Fig.~\ref{fig:2D_Dirac}\textbf{e},\textbf{h}.  Their transverse confinement lengths are all extremely similar and thus energy-independent, consistent with the previous straight-cut case; the same holds for mass distributions with nonzero minigaps. In Sec.~S4, we compare the DBC to the interface states of the Bernevig-Hughes-Zhang (BHZ) model \cite{Bernevig2006}, which is the continuum limit of the Wu-Hu model \cite{Wu2015,Bernevig2006}.  We show that these two families of 1D states are not equivalent, except for the limiting case of gapless ($\Delta\theta=\pi$) DBCs, which map to a subclass of BHZ interface states.  BHZ interface states do not support \textit{massive} relativistic dynamics, nor do they generally exhibit the DBC modes' energy-independent transverse confinement.



The Jackiw--Rossi model has previously been realized in several classical-wave experiments \cite{Kekule, Hou2007, Gutierrez2016, Gao2019, Noh2020, Gao2020, Menssen2020_braiding}, and although those works focused on phase singularities and the associated Jackiw--Rossi modes (also called ``Dirac vortex states'' in this context), we can leverage on such ideas to realize DBC modes.  We choose an acoustic crystal design based on the work of Torrent and S\'anchez-Dehesa \cite{Torrent2012}.  Starting from a triangular lattice of pillars of equal radius $r_0$ and next-nearest-neighbor distance $a$, we  apply the modulation shown in Fig.~\ref{fig:acoustic}\textbf{a}.  Each pillar is kept at a fixed center position $\mathbf{r}$, with radius
\begin{equation} \label{eq:Kekule_acoustic}
  R(\textbf{r}) = R_0 + \Delta R \,
  \mathrm{cos}\left[\textbf{K} \cdot \textbf{r} + \theta(\textbf{r})\right],
\end{equation}
\noindent
where $\textbf{K} = \textbf{K}_{+} - \textbf{K}_{-}$, with $\textbf{K}_{\pm}=\left[\pm 4 \pi/\sqrt{3}a,0 \right]$ denoting the corners of the original lattice's Brillouin zone \cite{Hou2007}.  The modulation couples the Dirac cones, and in the long wavelength limit the four mode amplitudes (two Dirac cones, with two amplitudes each) are governed by Eq.~\eqref{eq:hamiltonian} with $\Delta R \leftrightarrow |m|$ and $\theta(\textbf{r}) \leftrightarrow \mathrm{arg}(m)$\cite{Gao2019}.  In our experiments, we take $a=29$~mm. Unless stated otherwise, we set $R_0=0.2a$ (which yields a Dirac frequency of 10.4~kHz for the unmodulated lattice) and $\Delta R=0.13d$ (see Supplementary Information, Sec.~S5).

\begin{figure*}
  \centering
  \includegraphics[width=0.95\textwidth]{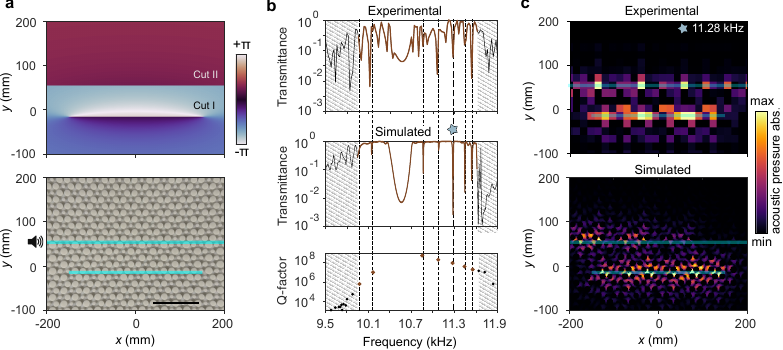}
  \caption{\textbf{Coupled waveguide-cavity system based on DBC modes.} \textbf{a}, Upper panel:  diagram $\theta (\textbf{r})$ created from branch function $f(z) = (z-150)^{1/2}(z+150)^{-1/2}$, with $z=x+i4y$. Lower panel: the acoustic crystal modulated with the diagram $\theta (\textbf{r})$. Blue dashed lines indicate the positions of cuts I and II, which are separated by a distance of $70$~mm. Scale bar: 100~mm. \textbf{b}, Upper panel: acoustic power transmittance to the right side of cut II. Middle panel: transmittance spectrum from FEM simulations. Lower panel: quality (Q) factors of the DBC modes on cut I. The bulk bands are shaded in gray. \textbf{c}, The profiles of DBC modes from experiment (upper panel) and simulations (lower panel) at the frequency marked by the star in \textbf{b}.}
  \label{fig:line_cavity}
\end{figure*}

We now implement a sample with a piecewise-constant phase distribution with $\Delta \theta =\pi$, analogous to Fig.~\ref{fig:2D_Dirac}\textbf{a}.  Fig.~\ref{fig:acoustic}\textbf{b} shows the phase distribution (left panel) and the modulated acoustic lattice (right panel).  We experimentally determine the excitation spectrum by applying an acoustic source and measuring the acoustic pressure at different points along the cut (Fig.~\ref{fig:acoustic}\textbf{c},\textbf{d}), followed by a Fourier transform (see Methods).  In Fig.~\ref{fig:acoustic}\textbf{c}, the color maps show the resulting excitation spectrum for this acoustic sample, as well as another similar sample with $\Delta \theta = 2\pi/3$.  Both sets of data agree quantitatively with the dispersion curves calculated directly from the acoustic equations via FEM (see Methods), plotted as cyan dashes.  These dispersion curves are approximately hyperbolic near the center of the bulk gap, in accordance with the theory, but deviate from the hyperbolic form as they approach the bulk bands.  For $\Delta\theta = \pi$, the DBC minigap vanishes and agrees with the prediction in Fig.~\ref{fig:2D_Dirac}\textbf{b}. For $\Delta\theta = 2\pi/3$, the minigap approximately matches the theoretical value of $2m_b$ calculated from Eq.~\eqref{dispersion} using the observed bulk gap.

We then measure the spatial profiles of the DBC modes in the frequency regime where the dispersion relation is hyperbolic. The results are plotted in Fig.~\ref{fig:acoustic}\textbf{e},\textbf{f} (left panel), and reveal exponential decays with distance $y$ from the cut.  In Fig.~\ref{fig:acoustic}\textbf{f} (left panel), we include the transverse profile predicted by the theoretical model (with $m_0$ again fitted using the observed bulk gap, and the Dirac velocity fitted to the slope of the dispersion curve in Fig.~\ref{fig:acoustic}\textbf{c}), which has decay length $\kappa^{-1} \approx 30$~mm.  Over the measured frequency range, which spans a large portion of the bulk gap, the experimental profiles agree well with the theoretical line. We also plot the exponential decays corresponding to the case with $\Delta\theta=2\pi/3$ in Fig.~\ref{fig:acoustic}\textbf{f} (right panel). These gapped DBC modes likewise retain nearly energy-independent transverse confinement over the investigated frequency range, and agree with the theoretical prediction.



To illustrate the flexibility of the DBC modes, we now study samples with different cut configurations.  In Fig.~\ref{fig:line_cavity}\textbf{a}, the upper panel shows a phase distribution constructed from a bounded cut (I) adjacent to an unbounded cut (II), forming a waveguide-cavity system like those found in integrated photonics \cite{Wang2021, Banerjee2025, Vahala2003}.  The DBC modes on cut I turn into a set of discrete modes for a line cavity (i.e., a Fabry-P\'erot-like resonator), while the DBC modes on cut II are waveguide modes that couple to the cavity.  A photograph of the experimental sample is shown in the lower panel of Fig.~\ref{fig:line_cavity}\textbf{a}.  We excite cut II from the left, and monitor the acoustic power transmitted to its other end.  Within the bulk gap, the measured transmittance spectrum (Fig.~\ref{fig:line_cavity}\textbf{b}, upper panel) exhibits a dip near mid-gap that may be attributed to local symmetry-breaking along the domain wall \cite{Xu2023} (as well as the reduction in group velocities above and below the minigap).  We also observe a sequence of narrow dips, which are absorption resonances caused by critical coupling to the cavity \cite{Vahala2003}.  These results are in agreement with the more ideal transmittance spectrum produced by FEM simulations (Fig.~\ref{fig:line_cavity}\textbf{b}, middle panel), including quantitative matches for the resonance frequencies (vertical dashes).  Furthermore, we numerically calculate the eigenmodes for the line cavity in the absence of cut II, and find that the eigenfrequencies closely match the absorption dips (Fig.~\ref{fig:line_cavity}\textbf{b}, bottom panel).  These simulation results also show that the eigenmodes' quality (Q) factors (based on radiative losses at the edges of the computational cell) remain approximately constant, or increase slightly, as the eigenfrequencies approach the upper bulk band.  In Fig.~\ref{fig:line_cavity}\textbf{c}, we present experimental measurements (upper panel) and simulations (lower panel) for the field intensity profile at one of the absorption resonances.  The excited line cavity mode is strongly confined to cut I, as expected.  

\begin{figure}
  \centering
  \includegraphics[width=0.45\textwidth]{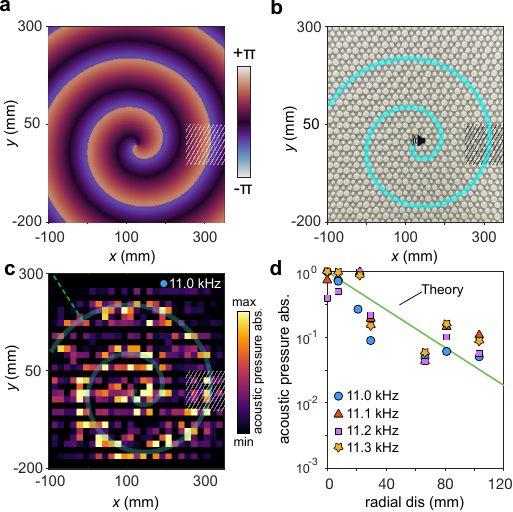}
  \caption{\textbf{DBC modes on a spiral.} \textbf{a}, A modulation phase profile $\theta (\textbf{r})$ with a spiral-shaped cut. \textbf{b}, Photograph of the acoustic lattice generated from this phase profile, with the cut indicated by the cyan curve. The location of the acoustic source in the experiment is indicated by the loudspeaker symbol. \textbf{c}, The acoustic pressure distribution measured at 11~kHz, a frequency within the potential barrier. \textbf{d}, Spatial profiles measured along a line transverse to the spiral (dashes in \textbf{c}) at four different frequencies (blue dots and red circles). In \textbf{a}--\textbf{c}, the shaded regions denote the regions with a potential barrier between $10.5$~kHz and $11.1$~kHz, assigned a modulation parameter of $R_0=0.17a$. The solid green line is the theoretical prediction.}
  \label{fig:spiral_waveguide}
\end{figure}

DBC modes have several other major advantages.  One is their inherent flexibility: they can transmit along arbitrary paths that can be easily generated by assigning branch-cuts.  Another advantage is that DBC modes support a form of Klein tunnelling \cite{Katsnelson2006, Jiang2020, Yu2024} due to their effective Dirac description; in the Supplementary Information, Sec.~S6, we show that the DBC mode along a straight waveguide can propagate through a potential barrier with insignificant reflection.

To demonstrate the utility of these two features, we design a spiral-shaped cut with $\Delta\theta = \pi$, as shown in Fig.~\ref{fig:spiral_waveguide}\textbf{a}. Keeping $|m|$ constant, we use the same scheme as before to generate the acoustic structure. We locally introduce a potential barrier to the spiral path (see the shaded region in Fig.~\ref{fig:spiral_waveguide}\textbf{a}). The potential barrier spans the frequency range from $10.5$~kHz to $11.1$~kHz and is realized by adding an offset of $-0.03a$ to the modulation parameter $R_0$. In the Supplementary Information, Sec.~S7, we show that the spiral waveguide exhibits high transmittance in a frequency range corresponding to the upper branch of DBC modes (see Fig.~\ref{fig:acoustic}\textbf{c}, left panel). The fabricated sample is shown in Fig.~\ref{fig:spiral_waveguide}\textbf{b}.  Placing an acoustic source at the center of the spiral, we probe the acoustic pressure distribution at regularly-spaced points across the sample. Evidently, the acoustic waves emitted by the source propagate outward along the spiral and bypass the potential barrier (Fig.~\ref{fig:spiral_waveguide}\textbf{c}), with no sign of Anderson localization in the longitudinal direction.  Fig.~\ref{fig:spiral_waveguide}\textbf{d} shows the field profiles along a line perpendicular to the spiral; the results measured at four representative frequencies exhibit confinement lengths consistent with the continuum theory (solid green line). These results show that the acoustic lattice reproduces the key features of the underlying relativistic continuum model, even along a strongly-deformed freeform cut.

In the Supplementary Information, Sec.~S7, we compare these results to the case of freeform waveguiding along a spiral path using conventional defect modes. The DBC modes provide significantly better mean transmittance and transmission bandwidth. Unlike a conventional interface between two globally distinct domains, a cut followed by a DBC can terminate at a branch point. This makes possible the present spiral, for which the regions on each side of the cut are globally connected. Comparing to freeform topological waveguides generated using amorphous lattices \cite{Wang2021, Banerjee2025}, DBC waveguides are at least as performant, but are easier to design since only a $\theta(\mathbf{r})$ profile is required (i.e., there is no need to generate an amorphous lattice by computational close-packing). 

In Supplementary Information, Sec.~S8, we numerically demonstrate the extension of the DBC design to photonic crystals and highlight its potential for higher-frequency operation through geometric scaling.


In summary, we have identified a previously unrecognized family of propagating Dirac defect states confined to branch cuts in a 2D complex Dirac mass field. These Dirac branch-cut (DBC) modes are distinct from the Jackiw--Rebbi \cite{Jackiw1976} and Jackiw--Rossi \cite{Jackiw1981} states, which are respectively bound to domain wall and vortex defects, and from the corner and hinge states associated with higher-order topological insulators \cite{Song2017PRL, Benalcazar2017PRB, Peterson2018, Noh2018}. The DBC modes propagate relativistically along the 1D cut, and can be massless or massive.  The key properties predicted by the continuum theory are replicated in a discrete acoustic lattice, including relativistic (linear or hyperbolic) dispersion, energy-independent transverse confinement, Klein tunnelling, and transport along strongly deformed freeform trajectories. We have also demonstrated that the DBC modes can form line-segment cavities, and can propagate along complicated spiral trajectories. In future, we expect that the lattice modulation scheme can be further refined to bring the actual DBC modes closer to their ideal counterparts, including making the dispersion relations hyperbolic over a larger portion of the bulk gap. Thus, our results establish branch-cut defects as a platform for investigating relativistic Dirac physics on lower-dimensional defects and for developing versatile wave-based device applications.

\section*{Methods}

\subsection*{Solution for a straight phase dislocation}

Starting from the Hamiltonian Eq.~\eqref{eq:hamiltonian}, we let the mass distribution be invariant along $x$, i.e., $m_1 + i m_2 = m(y)$.  We aim to derive the boundary states for a phase dislocation along the line $y = 0$.  Introducing the wavenumber $k_x$, the Schr\"odinger equation reduces to
\begin{align}
  \left[\tau_3 \left(\sigma_1 k_x - i \sigma_2 \partial_y\right)
    + m_0(y)\, U_{\theta}^\dagger \tau_1 U_{\theta} \right] \psi
  = E \, \psi,
  \label{eq:hamk_si}
\end{align}
where $U_\theta \equiv e^{i\theta\tau_3/2}$, and $\psi(y)$ is a four-component wavefunction.
We now define
\begin{equation}
  \psi(y) = U^\dagger_{\theta(y)} \, \varphi(y).
  \label{eq:varphi_def}
\end{equation}
When this is plugged into Eq.~\eqref{eq:hamk_si},
the $\partial_y$ gives rise to an additional gauge term because $\theta$ depends implicitly on $y$:
\begin{align}
  \partial_y \left[U^\dagger_{\theta} \varphi(y)\right]
  &= U^\dagger_{\theta} \left[\partial_y - \frac{i\theta'}{2} \,
    \tau_3\right] \, \varphi(y),
\end{align}
where $\theta' \equiv d\theta/dy$.  Hence, Eq.~\eqref{eq:hamk_si} simplifies to
\begin{equation}
  \left[
    \tau_3\sigma_1 k_x
    - i \tau_3\sigma_2 \partial_y
    - \frac{1}{2} \sigma_2\, \theta'
  + m_0(y) \tau_1 \right]
  \varphi = E \varphi.
  \label{eq:1dfield}
\end{equation}
Let us now choose the mass function
\begin{equation}
  m(y) = m_0 \, e^{iv_\theta y} \, \times \begin{cases} e^{i\theta_+}, & y > 0 \\
    e^{i\theta_-}, & y < 0,
  \end{cases}
  \label{eq:m_form}
\end{equation}
where $m_0 \in \mathbb{R}^+$ and $v_\theta \in \mathbb{R}$.  This has a phase discontinuity of $\Delta \theta = \theta_+ - \theta_-$ at $y = 0$.  The parameter $v_\theta$ gives a uniform phase gradient, with $v_\theta = 0$ corresponding to the piecewise-constant mass function in Fig.~1\textbf{a}.  Hence,
\begin{equation}
  \theta' = v_\theta + \Delta \theta\, \delta(y).
  \label{eq:gaugefield}
\end{equation}
We now take the ansatz
\begin{equation}
  \varphi(y) = \begin{cases} |\chi_+\rangle \, e^{-\kappa y}, & y > 0 \\
    |\chi_-\rangle \, e^{\kappa y}, & y < 0,
  \end{cases}
  \label{eq:chidef}
\end{equation}
where $\kappa \in \mathbb{R}^+$ and $|\chi_\pm\rangle$ are four-component spinors.  Plugging this into Eq.~\eqref{eq:1dfield}, for $y \ne 0$, gives
\begin{align}
  A_\pm |\chi_\pm\rangle &= E |\chi_\pm\rangle, \label{eq:chi1} \\
  A_\pm &\equiv  k_x \tau_3\sigma_1 \pm i \tau_3\sigma_2 \kappa
  - \frac{v_\theta}{2} \, \sigma_2 + m_0\tau_1.
  \label{eq:chi2} 
\end{align}

To handle the boundary conditions at $y = 0$, we require the \textit{original} wavefunction $\psi(y)$ to be continuous at $y = 0$, so the rotated wavefunction $\varphi(y)$ is discontinuous. Using Eq.~(\ref{eq:varphi_def}), the continuity of $\psi(y)$ requires $U_{\theta_+}^{\dagger}\varphi(0^+) = U_{\theta_-}^{\dagger}\varphi(0^-)$, which gives
\begin{equation}
\varphi(0^+)=U_{\Delta\theta}\varphi(0^-).
\label{eq:BC}
\end{equation}


Referring to Eq.~\eqref{eq:chidef}, this is satisfied by the boundary condition $|\chi_+\rangle = U_{\Delta \theta} \; |\chi_-\rangle$.  We can plug this into Eq.~\eqref{eq:chi1} to eliminate (say) $\chi_+$, obtaining
\begin{align}
  \left(U_{\Delta\theta}^\dagger A_+ U_{\Delta \theta} - A_-\right) |\chi_-\rangle = 0.
  \label{eq:UAU}
\end{align}
Using Eq.~\eqref{eq:chi2}, the terms in parentheses can be simplified to
\begin{align}
  2i\kappa\tau_3\sigma_2
  + m_0 \left[\tau_1 (\cos\Delta\theta-1) + \tau_2 \sin\Delta\theta \right].
     \label{eq:UAU_2}
\end{align}
What is the spectrum of this operator?  Pick an eigenstate of $\sigma_2$, i.e., $|\chi_-\rangle = |u\rangle\otimes |v\rangle$ where $\sigma_2 |v\rangle = \pm |v\rangle$.  Then Eq.~\eqref{eq:UAU} reduces to
\begin{align}
  \begin{pmatrix} \pm 2i\kappa &
    m_0\left(e^{-i\Delta\theta}-1\right) \\ m_0\left(e^{i\Delta\theta}-1\right)
    & \mp 2i\kappa
  \end{pmatrix} |u\rangle &= 0.
  \label{eq:chiral_choice}
\end{align}
For either choice of $\pm$, the determinant vanishes when
\begin{align}
  4\kappa^2 - m_0^2\; \Big|\cos\Delta\theta-1 -i\sin\Delta\theta\Big|^2
  = 0.
\end{align}
Hence, we arrive at Eq.~\eqref{kappa_fixed}.

Note that the result for $\kappa$ solves Eq.~\eqref{eq:chiral_choice} for both choices of $\pm$, i.e., both eigenstates of $\sigma_2$.  It follows that $|\chi_-\rangle$ is a superposition of these two choices.  To be specific, let us take the eigenstates of $\sigma_2$ to be $|v_+\rangle = 2^{-1/2} [1, i]^T$ and $|v_-\rangle = \sigma_1 |v_+\rangle$.  In each case, the non-Hermitian matrix in Eq.~\eqref{eq:chiral_choice} has equal and opposite eigenvalues that are either purely real or imaginary.  The zero eigenvalue condition is an exceptional point where the matrix has only one eigenvector, which we denote by $|u_+\rangle$ or $|u_-\rangle$.  Evidently, $|u_+\rangle$ and $|u_-\rangle$ depend only on $\Delta\theta$ (and are independent of $m_0$ and $k_x$), and can be defined such that
\begin{equation}
  |u_+\rangle = \frac{1}{\sqrt{2}} \begin{pmatrix}
    \textrm{sgn}[\sin(\Delta\theta/2)] \\
    e^{i\Delta\theta/2}
  \end{pmatrix}, \quad
  |u_-\rangle = \tau_3 |u_+\rangle.
  \label{eq:uvectors}
\end{equation}
Note that $\langle u_+ | u_- \rangle = 0$.  With these, the eigenproblem for $|\chi_-\rangle$ is converted into a restricted eigenproblem (i.e., one where the solution is constrained to a subspace):
\begin{align}
  \left[ k_x \tau_3\sigma_1 - i \tau_3\sigma_2 \kappa
    - \frac{v_\theta}{2}\sigma_2
    + m_0\tau_1\right]&
  |\chi_-\rangle = E |\chi_-\rangle,
  \label{eq:restricted_evproblem}
\end{align}
where $\kappa$ is given by Eq.~\eqref{kappa_fixed} and
\begin{align}
  |\chi_-\rangle &= \alpha \, |u_+\rangle\, |v_+\rangle + \beta \, |u_-\rangle \, |v_-\rangle
\end{align}
for some $\alpha, \beta \in \mathbb{C}$.  Applying either $\langle u_+|\, \langle v_+|$ or $\langle u_-|\, \langle v_-|$ to the left side, we obtain a pair of scalar equations for $\alpha$ and $\beta$ that can be consolidated into
\begin{align}
  &\Big(m_b \, \sigma_3 + k_x \sigma_1
  \Big) \begin{pmatrix}\alpha \\ \beta
  \end{pmatrix} = E \begin{pmatrix}\alpha \\ \beta \end{pmatrix},
  \label{eq:effective_1dDirac} \\
  &m_b = m_0 \; \textrm{sgn}[\sin\Delta\theta] \, \Big|\cos(\Delta \theta/2)\Big| - \frac{v_\theta}{2}.
\end{align}
This 1D Dirac equation, with effective Dirac mass $m_b$, gives the dispersion relation Eq.~\eqref{dispersion}.  The spectral theorem now implies that the restricted eigenproblem Eq.~\eqref{eq:restricted_evproblem} has two solutions for each $k_x$, which is significant as restricted eigenproblems are not guaranteed to possess solutions.  This clarifies why the $4\times 4$ Dirac equation exhibits two, not four, boundary states for each $k_x$, and retroactively justifies the ansatz Eq.~\eqref{eq:chidef}.

\section*{Numerical solutions}

The results presented in Fig.~\ref{fig:2D_Dirac}\textbf{c}--\textbf{h} are obtained by numerically solving the Jackiw--Rossi equations using the FEM software package COMSOL Multiphysics with custom-defined partial differential equations expressed in weak form (see Supplementary Information).  In these calculations, the cut is located along the curve $y = A\mathrm{sin} [N\pi(x-z_1)/(z_2 - z_1)]$, where $z_1 = -3$, $z_2 = 3$, and $A=(z_2 - z_1)/6$.  We take $N=2$ in Figs.~\ref{fig:2D_Dirac}\textbf{c}--\textbf{e}, and $N=1$ in Figs.~\ref{fig:2D_Dirac}\textbf{f}--\textbf{h}.

The numerical simulations of the acoustic crystal, such as those shown in Fig.~\ref{fig:acoustic}\textbf{c} and Fig.~\ref{fig:line_cavity}\textbf{b}--\textbf{c}, are also done using COMSOL Multiphysics.  The simulations solve the acoustic equations outside the pillars, using air density 1.18 $\mathrm{kg}\,\mathrm{m}^{-3}$ and speed of sound 343 $\mathrm{ms}^{-1}$; the pillar surfaces are treated as hard boundaries, which is valid due to the large impedance difference between air and the resin material.  Due to the limited acoustic pressure in our experiment, nonlinearities are neglected \cite{He2016_QSH}.

When calculating the dispersion relations for the DBC modes (Fig.~\ref{fig:acoustic}\textbf{c}), we use a supercell with periodic and radiation conditions on the boundaries parallel to $y$ and $x$, respectively.  When calculating the transmitted power and mode profiles in Figs.~\ref{fig:line_cavity} and \ref{fig:spiral_waveguide}, radiation conditions are applied on all edges of the computational cell.  We also use the numerical model to calculate the band structure of the acoustic structure, and validate that the Dirac mode amplitudes are governed by the Jackiw--Rossi model as described in Ref.~\cite{Torrent2012}; for details, see the Supplementary Information.

\section*{Sample fabrication and characterization}

Each acoustic sample is composed of an upper plate, lower plate, and sealing plug. The upper plate is made of transparent acrylic with modulus 3000 MPa and density $1190\, \mathrm{kg}/\mathrm{m}^3$, with holes at designated positions for inserting acoustic probes.  The lower plate and sealing plug are manufactured via stereolithography, using photosensitive resin with modulus 2880 MPa and density 1100 $\mathrm{kg}/\mathrm{m}^3$.  The pillars are fabricated directly on the lower plate.

For the dispersion measurements of Fig.~\ref{fig:acoustic}\textbf{c}, we set up an acoustic source (Br\"uel $\&$ Kj\ae r Type 2735) to emit a monochromatic signal through a narrow tube (radius $\approx 1.5\,\textrm{mm}$ and length $100\,\textrm{mm}$) to a point on the sample's left edge.  The signal is picked up by two probe microphones (Br\"uel $\&$ Kj\ae r  Type 4182).  The first microphone scans over 48 small holes (diameter $2\,\textrm{mm}$, step size $12.56\,\textrm{mm}$) placed along the line of the phase dislocation (parallel to the $x$ direction).  During each measurement, the other holes are filled with sealing plugs to prevent acoustic leakage.  The second microphone is fixed towards the source to pick up a reference signal.  The acoustic pressure at different positions is recorded with a data collector (Br\"uel$\&$Kj\ae r  Type 3560-C), and a Fourier transform is performed to obtain the $k_x$-dependence.  This process is repeated at different frequencies to obtain the results in Fig.~\ref{fig:acoustic}\textbf{c}.

For the intensity profile measurements of Figs.~\ref{fig:line_cavity} and \ref{fig:spiral_waveguide}, we use a monochromatic acoustic source and two probe microphones in a similar manner.  The acoustic signal is picked up along a grid of holes with spacing $16.74\,\textrm{mm}$ in both the $x$ and $y$ directions.

\section*{Acknowledgements}

This work was supported by the Singapore National Research Foundation (NRF) under Competitive Research Program (CRP) Nos.~NRF-CRP23-2019-0005, NRF-CRP23-2019-0007, and NRF-CRP29-2022-0003, and the NRF Investigatorship NRF-NRFI08-2022-0001; the Singapore Ministry of Education Singapore (MOE-T2EP50122-0019); the Singapore A*STAR Grant No.~A2090b0144; the China Scholarship Council: 202406280469.

B.Z. and Y.D.C. conceived the idea. B.Z. performed the theoretical derivations with the help from Y.D.C. B.Z. performed the numerical calculations and designed the experiments. C.M. conducted the measurements with the help from X.H.Z.. B.Z. analysed the experimental data with the help from C.M. B.Z., Y.D.C. and B.L.Z. wrote the manuscript. Q.W., G.G.L. and Z.C. discussed and commented on the manuscript. Q.J.W., B.L.Z. and Y.D.C. supervised the project.

\clearpage

\begin{widetext}
\makeatletter 
\renewcommand{\theequation}{S\arabic{equation}}
\makeatother
\setcounter{equation}{0}

\makeatletter 
\renewcommand{\thesection}{S\arabic{section}} 
\setcounter{section}{0}

\makeatletter 
\renewcommand{\thefigure}{S\@arabic\c@figure}
\makeatother
\setcounter{figure}{0}

\makeatletter 
\renewcommand{\thetable}{S\@arabic\c@table}
\makeatother
\setcounter{table}{0}

\begin{center}
  {\large \textbf{Supplementary Information for}}\\
  \vskip 0.05in
  {\Large ``Dirac branch-cut modes with relativistic transport''}\\
  \vskip 0.1in

  Bofeng Zhu, Chengzhi Ma, Qiang Wang, Gui-Geng Liu, \\
  Xiuhai Zhang, Zheyu Cheng, Qi Jie Wang, Baile Zhang, and Y.~D.~Chong
\end{center}

\section{Scattering matrix calculations for straight phase dislocations}

In this section, we describe a direct numerical solution for the Jackiw-Rossi model for a straight semi-infinite phase dislocation, which is helpful for checking the analytical derivations (see Fig.~1\textbf{b} in the main text).  Starting from the $4\times4$ Dirac equation in 2D, we replace $-i\partial/\partial x \rightarrow k_x$ and write the Hamiltonian as
\begin{equation} \label{eq:si_Hmatrix}
  H = 
 \begin{bmatrix}
 0 & k_x - \partial_y & m(y) & 0 \\ 
 k_x + \partial_y & 0 & 0 & m(y) \\
 m^*(y) & 0 & 0 & -k_x + \partial_y \\
 0 & m^*(y) & -k_x - \partial_y & 0 \\
 \end{bmatrix}.
\end{equation}
\noindent
For simplicity, we assume the piecewise-constant $m$ function shown in Fig.~\ref{fig:2D_Dirac}\textbf{a} of the main text (i.e., $v_\theta = 0$ and $\theta_- = 0$). Then \eqref{eq:si_Hmatrix} can be written as
\begin{align}
  H &= H_0(m,k_x) + \Lambda \frac{\partial}{\partial y},
  \label{eq:H_decomp1} \\
  H_0(m,k_x) &= 
  \begin{bmatrix}
    0 & k_x & m & 0 \\ 
    k_x & 0 & 0 & m \\
    m^* & 0 & 0 & -k_x \\
    0 & m^* & -k_x & 0 \\
  \end{bmatrix}, \;\;\;
  \Lambda = \begin{bmatrix}
    0 & -1 & 0 & 0 \\ 
    +1 & 0 & 0 & 0 \\
    0 & 0 & 0 & +1\\
    0 & 0 & -1 & 0 \\
  \end{bmatrix} = - \Lambda^{-1}.
  \label{eq:H_decomp2}
\end{align}

First, consider $y \le 0$.  Taking $\theta_- = 0$ for simplicity (i.e., $m = m_0$), we seek solutions for fixed $E$ of the form
\begin{align}
  \psi(y) = e^{\gamma y} |\mu\rangle, \quad
  H \psi(y) = E \psi(y).
\end{align}
Plugging this into \eqref{eq:H_decomp1}--\eqref{eq:H_decomp2}, and doing some rearrangement, yields
\begin{equation}
  \Lambda \Big[H_0(m_0, k_x) - E\Big] |\mu\rangle = \gamma |\mu\rangle.
  \label{eq:evanescent_eigenproblem}
\end{equation}
It can be shown that this eigenproblem's eigenvalues come in pairs $\{\pm\gamma_1, \pm \gamma_2\}$, which can be either real (evanescent waves) or imaginary (propagating waves).  We use the $+$ label for evanescent waves that grow to the right ($\gamma \in \mathbb{R}^+$) as well as propagating waves with positive phase velocity ($\gamma = ik$ where $k \in \mathbb{R}^+$); and conversely for the $-$ label.  The corresponding eigenvectors are denoted by $\{|\mu_1^\pm\rangle, |\mu_2^\pm\rangle$.  Hence, the wavefunction has the form
\begin{equation}
  \psi(y)
  = a_1 e^{\gamma_1 y} |\mu_1^+\rangle
  + a_2 e^{\gamma_2 y} |\mu_2^+\rangle
  + b_1 e^{-\gamma_1 y} |\mu_1^-\rangle
  + b_2 e^{-\gamma_2 y} |\mu_2^-\rangle, \;\;\;(y \le 0),
  \label{eq:psi_boundary1}
\end{equation}
for some coefficients $a_{1(2)}, b_{1(2)} \in \mathbb{C}$.

The procedure for $y \ge 0$ is similar, but with $m = m_0e^{i\Delta\theta}$.  After constructing the eigenproblem analogous to \eqref{eq:evanescent_eigenproblem}, we obtain a second set of eigenvalues $\{ \pm \eta_1 , \pm \eta_2\}$, with eigenvectors $\{|\nu_1^\pm\rangle, |\nu_2^\pm\rangle\}$.  The wavefunction has the form
\begin{equation}
  \psi(y)
  = c_1 e^{\eta_1 y} |\nu_1^+\rangle
  + c_2 e^{\eta_2 y} |\nu_2^+\rangle
  + d_1 e^{-\eta_1 y} |\nu_1^-\rangle
  + d_2 e^{-\eta_2 y} |\nu_2^-\rangle, \;\;\;(y \ge 0),
  \label{eq:psi_boundary2}
\end{equation}
for $c_{1,2}, d_{1,2}\in \mathbb{C}$.  We now impose continuity of the four-component wavefunction, by equating \eqref{eq:psi_boundary1} to \eqref{eq:psi_boundary2} at $y = 0$:
\begin{align}
  a_1 |\mu_1^+\rangle + a_2 |\mu_2^+\rangle + b_1 |\mu_1^-\rangle
  + b_2 |\mu_2^-\rangle &=
  c_1 |\nu_1^+\rangle + c_2 |\nu_2^+\rangle + d_1  |\nu_1^-\rangle
  + d_2  |\nu_2^-\rangle \\
  \Rightarrow
  a_1 |\mu_1^+\rangle + a_2 |\mu_2^+\rangle
  - d_1  |\nu_1^-\rangle - d_2  |\nu_2^-\rangle
  &= - b_1 |\mu_1^-\rangle - b_2 |\mu_2^-\rangle
  + c_1 |\nu_1^+\rangle + c_2 |\nu_2^+\rangle \\
  \Rightarrow\;\;\;\,
  \Bigg[
    |\mu_1^+\rangle, |\mu_2^+\rangle,
    - |\nu_1^-\rangle, -|\nu_2^-\rangle
    \Bigg]
  \begin{bmatrix} a_1 \\ a_2 \\ d_1 \\ d_2
  \end{bmatrix} &=
  \Bigg[
    - |\mu_1^-\rangle, -|\mu_2^-\rangle,
    |\nu_1^+\rangle, |\nu_2^+\rangle
    \Bigg]
  \begin{bmatrix} b_1 \\ b_2 \\ c_1 \\ c_2
  \end{bmatrix}.
\end{align}
On the left, we group together the amplitudes for the modes growing/propagating \textit{toward} the domain wall, and on the right we have the modes growing/propagating \textit{away from} the domain wall.  Inverting the matrix on the right, we arrive at a scattering matrix relation of the form
\begin{equation} \label{eq:10}
 S(k_x, E, m_0, \Delta \theta)  \begin{bmatrix}
 a_1 \\ a_2 \\ d_1 \\ d_2 \\
 \end{bmatrix} = 
 \begin{bmatrix}
 b_1 \\ b_2 \\ c_1 \\ c_2 \\
 \end{bmatrix}.
\end{equation}
Now we can vary $E$ and/or other parameters, and monitor the eigenvalues of $S$.  A zero eigenvalue corresponds to the existence of an eigenstate of $H$ that is exponentially localized to $y = 0$.  (Note that propagating waves cannot yield zero eigenvalues due to energy conservation).  In Fig.~1\textbf{b} of the main text, we plot the minimum squared eigenvalue, $\mathrm{min}[\mathrm{eig}(S^{\dagger}S)]$, versus $E$ and $k_x$ and/or $\Delta\theta$, and use the zeros of this quantity to locate the Jackiw-Rebbi-like bound states, which agree well with the previous analytical results.

\section{Numerical solutions of bound states at 2D phase dislocation}
\label{si:num_pde}

To solve the Jackiw--Rossi equations using the finite element method (FEM), we express the equations in weak form.  We start with the time-independent Schr\"odinger equation
\begin{equation} \label{eq:Jackiw-Rossi}
 \begin{bmatrix}
 0 & -i \partial_x - \partial_y & m_1-im_2 & 0 \\ 
 -i \partial_x + \partial_y & 0 & 0 & m_1-im_2 \\
 m_1+im_2 & 0 & 0 & i\partial_x + \partial_y \\
 0 & m_1+im_2 & i\partial_x - \partial_y & 0 \\
 \end{bmatrix}
 \begin{bmatrix}
   \psi_1 \\ \psi_2 \\ \psi_3  \\ \psi_4
 \end{bmatrix}
 = E \begin{bmatrix}
   \psi_1 \\ \psi_2 \\ \psi_3  \\ \psi_4
 \end{bmatrix},
\end{equation}
where $\psi_{1,2,3,4}$ are the wavefunction components and $E$ is the energy eigenvalue.  Next, we multiply each component of this equation with a set of test functions $\phi_{1,2,3,4}$, and integrate over some area $\Omega$:
\begin{align} \label{eq:weak1}
\int_{\Omega} \phi_1 \cdot \big[-i (\partial_x \psi_2 - i\partial_y \psi_2) + (m_1 - im_2) \psi_3\big] \; d^2r &= E \int_{\Omega}  \phi_1 \cdot \psi_1 \; d^2r \\
\int_{\Omega} \phi_2 \cdot \big[-i (\partial_x \psi_1 + i\partial_y \psi_1) + (m_1 - im_2) \psi_4\big] \; d^2r &= E \int_{\Omega}  \phi_2 \cdot \psi_2 \; d^2r \\
\int_{\Omega} \phi_3 \cdot \big[(m_1 + im_2) \psi_1 + i (\partial_x \psi_4 - i\partial_y \psi_4)\big] \; d^2r &= E \int_{\Omega} \phi_3 \cdot \psi_3 \;d^2r\\
\int_{\Omega} \phi_4 \cdot \big[(m_1 + im_2) \psi_2 + i (\partial_x \psi_3 + i\partial_y \psi_3)\big] \; d^2r &= E \int_{\Omega} \phi_4 \cdot \psi_4 \;d^2r.
\label{eq:weak4}
\end{align}
We then apply integration by parts to the derivative terms on the left side of each equation.  For example, for the first term in the left hand side of \eqref{eq:weak1}, we have
\begin{align}\label{eq:17}
  \int_{\Omega} \phi_1 \cdot (-i \partial_x \psi_2) \, d^2r
  = i \int_{\Omega} (\partial_x \phi_1) \cdot \psi_2 \, d^2r
  - i \oint_{\partial\Omega} \left(\phi_1 \cdot \psi_2\right) n_x \; d\ell,
\end{align}
where $\Omega$ and $\partial \Omega$ are the domain and its boundary, $n_x$ is the $x$-component of outward normal vector on $\partial \Omega$, and $d\ell$ is the boundary element.  The first term on the right of \eqref{eq:17} enters into our desired weak form of the PDE.  The second term is a boundary term that is eliminated in the usual way (e.g., by imposing Dirichlet conditions on distant boundaries).  Carrying out this procedure for all the terms in \eqref{eq:weak1}--\eqref{eq:weak4}, we obtain
\begin{align}\label{eq:8}
\int_{\Omega}  \big[+ i (\partial_x \phi_1) \psi_2 +  (\partial_y \phi_1) \psi_2 + (m_1 - im_2) \phi_1 \psi_3 - E \phi_1 \psi_1 \big] d^2r &= 0 \\
\int_{\Omega}  \big[+ i (\partial_x \phi_2) \psi_1 -  (\partial_y \phi_2) \psi_1 + (m_1 - im_2) \phi_2 \psi_4 - E \phi_2 \psi_2 \big] d^2r &= 0 \\
\int_{\Omega}  \big[- i (\partial_x \phi_3) \psi_4 -  (\partial_y \phi_3) \psi_4 + (m_1 + im_2) \phi_3 \psi_1 - E \phi_3 \psi_3 \big] d^2r &= 0 \\ 
\int_{\Omega}  \big[- i (\partial_x \phi_4) \psi_3 +  (\partial_y \phi_4) \psi_3 + (m_1 + im_2) \phi_4 \psi_2 - E \phi_4 \psi_4 \big] d^2r &= 0
\end{align}
The integrands are the weak expressions entered into the FEM solver.  The phase and amplitude of the complex mass are input as $m_1 = |m(\textbf{\textbf{r}})|\cos(\theta(\textbf{r}))$ and $m_2 = |m(\textbf{r})|\sin(\theta(\textbf{r}))$.  The numerical solutions are obtained using Dirichlet boundary conditions, $\psi_{1,2,3,4}=0$, at the edges of the computational cell, far from the plotted region.

\begin{figure}
  \centering
  \includegraphics[width=0.75\textwidth]{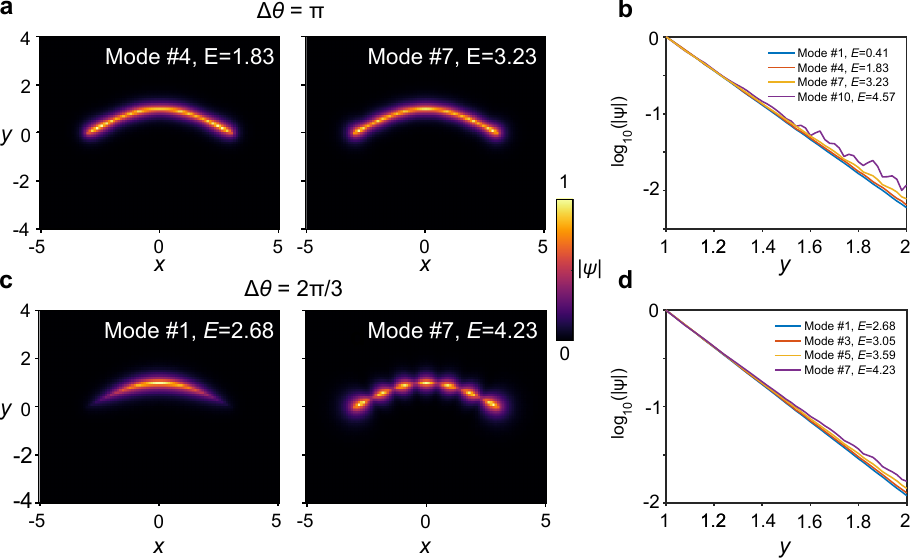}
  \caption{\textbf{DBC modes on a bounded cut}. \textbf{a}, Spatial profiles of DBC modes on a curved cut with $\Delta\theta = \pi$.  All the other parameters are the same as in Fig.~1\textbf{h} of the main text.  \textbf{b}, Profiles along the line $x=0$ for the two DBC modes in \textbf{a}, as well as the two modes plotted in Fig.~1\textbf{h} of the main text.  All four modes have very similar confinement lengths.  \textbf{c}--\textbf{d}, Mode profiles for $\Delta\theta = 2\pi/3$.}
  \label{fig:high_order_DBC modes}
\end{figure}

\section{Higher-order DBC modes}

In Fig.~1\textbf{h} of the main text, we plotted the spatial profiles for several Dirac branch-cut (DBC) modes on an exemplary bounded cut.  Here, we present the spatial profiles for several other modes.  In Fig.~\ref{fig:high_order_DBC modes}\textbf{a}, we plot the profiles for DBC modes \# 4 and \# 7 at $\Delta\theta = \pi$ (all other parameters are the same as in Fig.~1\textbf{h} of the main text).  It can be seen that the mode profiles are extremely similar to those of modes \# 1 and \# 10 from in the main text.  In Fig.~\ref{fig:high_order_DBC modes}\textbf{b}, we plot all four modes' profiles along the center line $x = 0$.  This reveals that the confinement lengths are all highly similar, consistent with the analytic prediction derived from straight cuts.

In Fig.~\ref{fig:high_order_DBC modes}\textbf{c}, we plot the spatial profiles of two representative modes for the $\Delta\theta = 2\pi/3$ case.  The corresponding profiles along $x = 0$ are plotted in Fig.~\ref{fig:high_order_DBC modes}\textbf{d}.  Once again, the modes at very different energies have the same confinement lengths, averaging to $\kappa \approx 0.87$, while the analytic value [Eq.~(5) of the main text] is $\kappa = \sin(\pi/3)/\sin(\pi/2) \approx 0.86603$.

\begin{figure}
  \centering
  \includegraphics[width=0.83\textwidth]{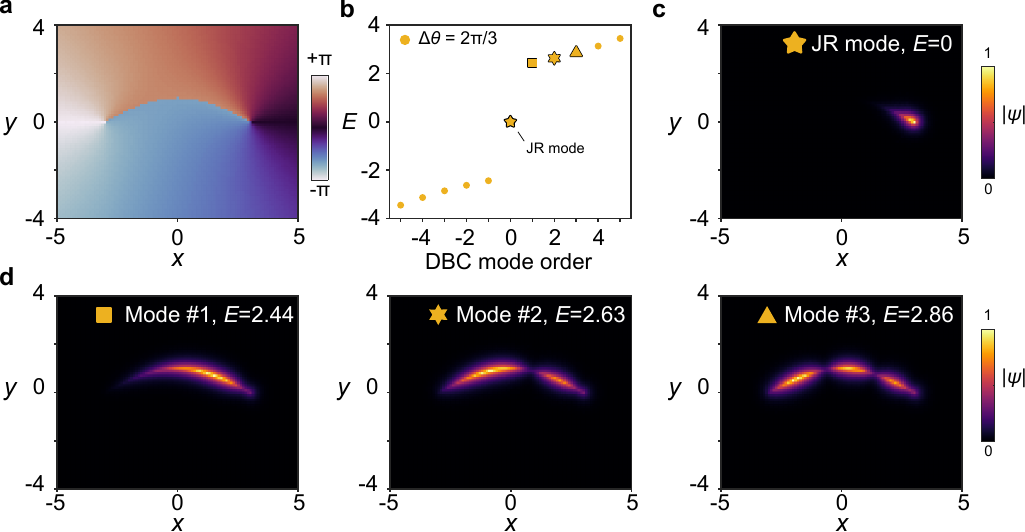}
  \caption{\textbf{Separation of Jackiw--Rossi states and DBC modes}.  \textbf{a}, Phase map for the mass field $m(\mathbf{r}) = (z-3)^{1/3}(z+3)^{2/3}$, using a sinusoidal half-period (with amplitude 1) for the cut.  \textbf{b}, Eigenenergy spectrum, showing the Jackiw--Rossi state at $E = 0$ and a sequence of DBC modes with a finite gap.  \textbf{c}--\textbf{d}, Spatial profiles for the Jackiw--Rossi mode, and three representative DBC modes indicated by the corresponding markers in \textbf{b}. }
  \label{fig:JR_gapped_DBC}
\end{figure}

In Fig.~1\textbf{c}--\textbf{e} of the main text, we presented a case where the Jackiw--Rossi state coincides with one of the DBC modes, and has properties similar to all the other DBC modes.  It is also possible to design $m(\mathbf{r})$ functions hosting Jackiw--Rossi modes distinct from DBC modes.  For example, Fig.~\ref{fig:JR_gapped_DBC}\textbf{a} shows the phase distribution for the mass field $m(\mathbf{r}) = (z-3)^{1/3}(z+3)^{2/3}$.  This has a net winding of $2\pi$ at large distances, and therefore hosts a Jackiw--Rossi state at $E = 0$.  At the same time, the phase discontinuity at the cut is $2\pi/3$, so the DBC modes are gapped---i.e., their energies cannot overlap with that of the Jackiw--Rossi state, as shown in Fig.~\ref{fig:JR_gapped_DBC}\textbf{b}.  The spatial profiles for several modes are plotted in Fig.~\ref{fig:JR_gapped_DBC}\textbf{c}--\textbf{d}.  In this case, we find that the Jackiw--Rossi mode does not spread out along the cut, as the DBC modes do, but is localized near one of the ends.

\section{Comparison to the Bernevig-Hughes-Zhang (BHZ) model}

In this section, we compare the DBC modes to the topological interface states of Bernevig-Hughes-Zhang (BHZ) type Hamiltonians \cite{Bernevig2006}, which constitute the continuum limit for Wu-Hu lattices \cite{Wu2015}. We begin with a block diagonal $k\cdot p$ type Hamiltonian in the ($p_{+}, d_{+}, p_{-}, d_{-}$) basis \cite{Wu2015},
\begin{equation} \label{eq:kp_Ham_full}
  H' = 
 \begin{bmatrix}
 H_{+} & 0 \\ 
 0 & H_{-} \\
 \end{bmatrix},  
 H_{\pm} = 
 \begin{bmatrix}
 m_0+Bk^2 & A^*k_{\pm}\\ 
 Ak_{\mp} & -m_0-Bk^2 \\
 \end{bmatrix},
\end{equation}
where $k_{\pm} = k_x \pm i k_y$ and $k^2 = k_x^2 + k_y^2$, $m_0 = (\epsilon_p^0 -  \epsilon_d^0)/2$ is the real Dirac mass, $B$ and $A=v i$ are positive real and purely imaginary numbers, respectively, and $v$ is the velocity. The bulk dispersion is
\begin{equation}\label{eq:kp_PDE_dispersion}
E_{\pm} = \pm \sqrt{(m_0+Bk^2)^2 + |A|^2k^2}.
\end{equation}

Let us take a straight interface parallel to $x$ so that $k_x$ is conserved. In one uniform half-space, the mode bounded at $y=0$ must have an evanescent solution $\Phi(x,y) \propto e^{ik_xx-\kappa y}, \kappa>0$. By inserting replacements $k_y \rightarrow i\kappa$ and $k^2 \rightarrow k_x^2 - \kappa^2$ into the bulk dispersion Eq.~(\ref{eq:kp_PDE_dispersion}), we have
\begin{equation}
E^2 = (m_0+B(k_x^2-\kappa^2))^2 + |A|^2(k_x^2-\kappa^2).
\end{equation}
After solving this quadratic equation, we find that the decay constant satisfies
\begin{equation}\label{eq:kp_PDE_dispersion_decay}
\kappa^2 = k_x^2 - q_{\pm}(E)
\end{equation}
where
\begin{equation}\label{eq:kp_PDE_dispersion_decay_q}
 q_{\pm}(E) = \frac{-(2m_0B + |A|^2) \pm \sqrt{|A|^4+4m_0B|A|^2 +4B^2E^2}}{2B^2}.
\end{equation}

In the $B=0$ limit, this Hamiltonian reduces to a first-order Dirac theory whose straight-interface state is equivalent to the $\Delta\theta=\pi$ DBC discussed in the main text. For $B\neq0$, Eqs.~(\ref{eq:kp_PDE_dispersion_decay})-(\ref{eq:kp_PDE_dispersion_decay_q}) therefore give an explicit dependence of the transverse decay constant on energy and longitudinal momentum, even if the dispersion is purely linear, as illustrated in Figs.~\ref{fig:WH_confinement}\textbf{a--b}.

Compare this to the DBC family.  Moving away from the shared point ($\Delta\theta=\pi$, see Figs.~\ref{fig:WH_confinement}\textbf{c--d}) opens a minigap, creating 1D \textit{massive relativistic} modes (see Figs.~\ref{fig:WH_confinement}\textbf{e--f}).  For a straight-line phase discontinuity, the Hamiltonian can be written as
\begin{equation}\label{DBC_Ham}
H_{\rm DBC} = v k_{\parallel}\alpha_{\parallel} -iv\alpha_{\perp}\partial_{\perp} +M\,\mathrm{sgn}(y)\beta_1+\mu\beta_2, 
\end{equation}
with
\begin{equation}\label{Eq3}
M=m_0\sin(\Delta\theta/2),\qquad \mu=m_0\cos(\Delta\theta/2),
\end{equation}
where $\beta_1$ and $\beta_2$ anticommute with each other and with the kinetic Dirac matrices.  Importantly, introducing the second mass does not generate higher-order momentum terms or destroy the Dirac algebra. Projection onto the confined subspace gives
\begin{equation}\label{Eq4}
h_{\rm DBC}=v k_{\parallel}\eta_z+\mu\eta_x,
\end{equation}
and hence
\begin{equation}\label{Eq5}
E^2=v^2k_{\parallel}^2+\mu^2, \qquad \kappa=|M|/v.
\end{equation}
In all cases, the transverse decay constant $\kappa=|M|/v$ is independent of longitudinal momentum and energy.

 \begin{figure}
   \centering
   \includegraphics[width=0.83\textwidth]{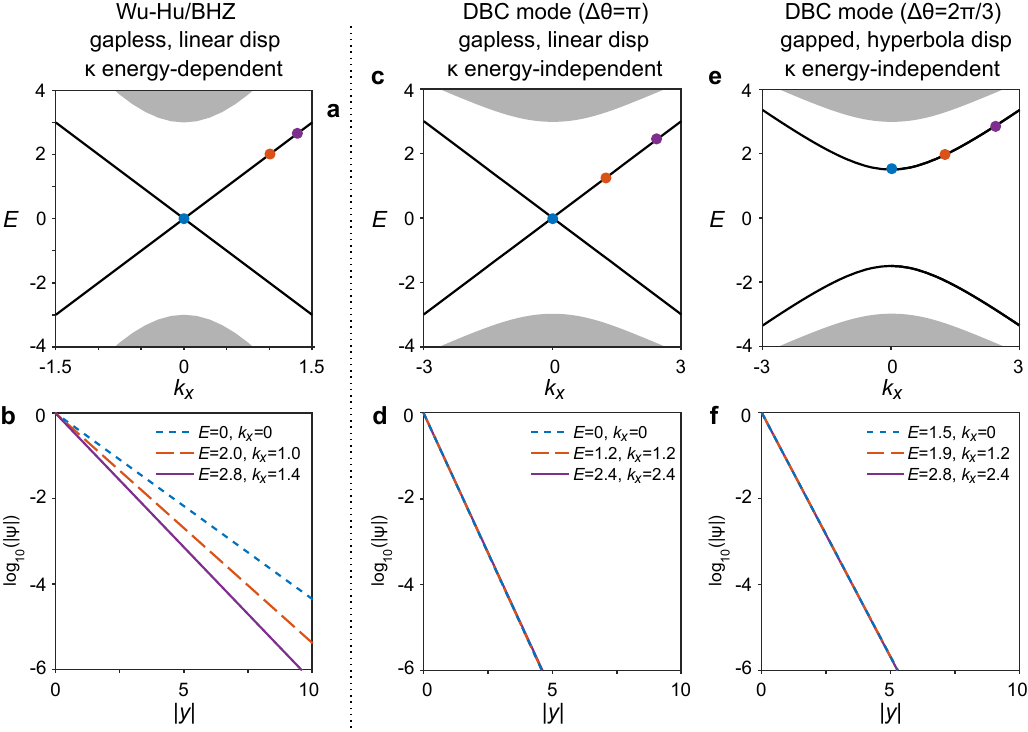}
   \caption{\textbf{Comparison of the energy dependence of transverse confinement in the Wu–Hu/BHZ model and DBC modes.} \textbf{a}, Wu-Hu/BHZ edge state dispersion. \textbf{b}, Transverse profile of the field strength $|\psi|$ for BHZ edge states into the positive-mass half-space, obtained at different energies $E$ and Bloch wavevector $k$ marked in \textbf{a}. \textbf{c}, DBC mode dispersion under a phase difference $\Delta\theta=\pi$. \textbf{d}, Transverse profile of the field strength $|\psi|$ for DBC modes marked in \textbf{c}. \textbf{e}, DBC mode dispersion under a phase difference $\Delta\theta=2\pi/3$. \textbf{f}, Transverse profile of the field strength $|\psi|$ for DBC modes marked in \textbf{e}. The Dirac mass is $m_0 = 3$. Other parameters in \textbf{a-b} are $B=1$ and $A=2i$.}
   \label{fig:WH_confinement}
 \end{figure}

\section{Acoustic structure design}
\label{sec:design}

\begin{figure}
  \centering
  \includegraphics[width=\textwidth]{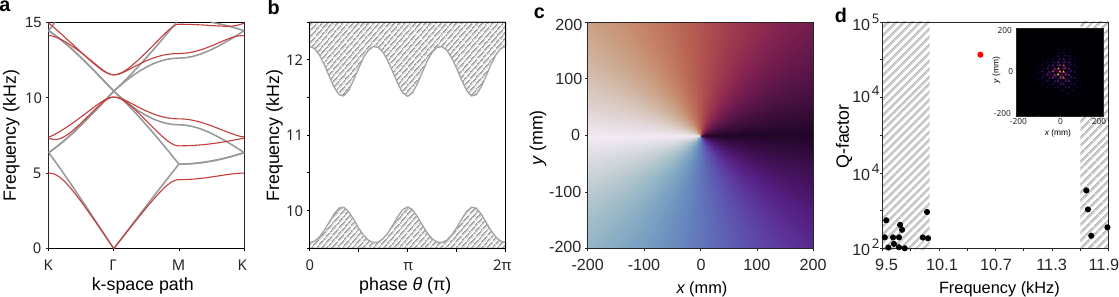}
  \caption{\textbf{Design of the acoustic structure.} \textbf{a}, Band structure for the acoustic crystal with Kekul\'e modulation (red lines, modulation phase $\theta=-0.35\pi$) and without it (gray lines).  All other parameters are the same as in Fig.~2 of the main text. \textbf{b} Bulk band gap versus modulation phase $\theta$, with bulk bands shaded in gray. \textbf{c}, A phase map $\theta(\mathbf{r})$ with a phase singularity. \textbf{d}, Eigenmode spectrum for the acoustic structure generated from \textbf{c}.  The state in the bulk gap is a Jackiw--Rossi or ``Dirac vortex'' state.  Inset: spatial profile of the in-gap state, which is localized at the center of the phase singularity and is thus interpretable as the Jackiw--Rossi mode \cite{Jackiw1981}.  All results shown here are obtained from FEM simulations.}
  \label{fig:acoustic_unit_cell}
\end{figure}

The acoustic structure that we use to realize the Jackiw--Rossi model, and the consequent DBC modes, is similar to the design in Ref.~\onlinecite{Torrent2012}.  We begin with a triangular lattice of solid pillars surrounded by air, as shown in Fig.~\ref{fig:acoustic}\textbf{a} of the main text.  If all pillars have equal radii, the band structure has two Dirac points at the corners of the hexagonal Brillouin zone.  If the lattice is viewed in terms of an expanded unit cell with six pillars each, both Dirac points are folded to the center, $\Gamma$, of the shrunken Brillouin zone \cite{Wu2015}.  In Fig.~\ref{fig:acoustic_unit_cell}\textbf{a}, the gray lines show the acoustic band structure calculated by FEM.  Near $\Gamma$, therefore, we have four bands (two from each valley) governed by an effective Jackiw--Rossi model with $m_1 = m_2 = 0$.

We then apply a Kekul\'e-type modulation by varying the pillar radii according to Eq.~(6) of the main text, which involves a modulation amplitude $\Delta R \ne 0$ and phase $\theta$.  The modulation lifts the Dirac point degeneracy, as seen in Fig.~\ref{fig:acoustic_unit_cell}\textbf{a} (the red lines show the bands for $\theta = -0.35\pi$).  Since the modulation phase parameter $\theta$ is an angle variable by construction, we infer that it maps to the phase parameter $\mathrm{arg}[m]$ for the complex mass field \cite{Gao2019, Gao2020}.  In fact, this parameter mapping is not exact: the modulation phase also induces a variation in $|m|$, as seen in the variation in the bulk gap in Fig.~\ref{fig:acoustic_unit_cell}\textbf{b}.  In the present work, we neglect the $|m|$ variation when designing our DBC waveguide structures; in the future, further optimizing the modulation scheme may enable creating more performant structures.

Thus, to realize a Jackiw--Rossi model with spatially-varying $\mathrm{arg}[m(\mathbf{r})]$, we apply the corresponding spatially-varying modulation phase parameter $\theta(\mathbf{r})$ \cite{Gao2019, Gao2020}.  For instance, it is possible to generate a Jackiw--Rossi (or ``Dirac vortex'') state with a $\theta(\mathbf{r})$ distribution like the one shown in Fig.~\ref{fig:acoustic_unit_cell}\textbf{c}, containing a phase singularity.  Using this $\theta(\mathbf{r})$, we generate the acoustic structure and calculate its eigenmodes; the results are shown in Fig.~\ref{fig:acoustic_unit_cell}\textbf{d}, and indeed feature a localized eigenmode in the gap.  In the main text, we proceed to study $\theta(\mathbf{r})$ distributions containing branch-cuts, which host DBC modes.
 
\section{Klein tunnelling of DBC modes}

In this section, we demonstrate Klein tunnelling for the $\Delta\theta=\pi$ DBC mode used in the freeform-waveguiding experiment. Klein tunnelling follows from the effective Dirac dynamics and allows transmission through a scalar potential barrier with strongly suppressed reflection. This phenomenon has been previously demonstrated for massless Dirac particles in graphene and photonic crystal \cite{Katsnelson2006,Jiang2020} and massive Dirac particles in time-domain synthetic lattice \cite{Yu2024}.

We first verify Klein tunnelling in the continuum model by numerically solving a set of weak form PDEs modified from the Eq.~\ref{eq:8} in Sec.~\ref{si:num_pde}.

\begin{align}\label{eq:transport_pde}
\int_{\Omega}  \big[+ i (\partial_x \phi_1) \psi_2 +  (\partial_y \phi_1) \psi_2 + (m_1 - im_2) \phi_1 \psi_3 + (U - i\eta - E) \phi_1 \psi_1 - s_1 \phi_1 \big] d^2r &= 0 \\
\int_{\Omega}  \big[+ i (\partial_x \phi_2) \psi_1 -  (\partial_y \phi_2) \psi_1 + (m_1 - im_2) \phi_2 \psi_4 + (U - i\eta - E) \phi_2 \psi_2 - s_2 \phi_2 \big] d^2r &= 0 \\
\int_{\Omega}  \big[- i (\partial_x \phi_3) \psi_4 -  (\partial_y \phi_3) \psi_4 + (m_1 + im_2) \phi_3 \psi_1 + (U - i\eta - E) \phi_3 \psi_3  - s_3 \phi_3 \big] d^2r &= 0 \\ 
\int_{\Omega}  \big[- i (\partial_x \phi_4) \psi_3 +  (\partial_y \phi_4) \psi_3 + (m_1 + im_2) \phi_4 \psi_2 + (U - i\eta - E) \phi_4 \psi_4  - s_4 \phi_4 \big] d^2r &= 0
\end{align}
\noindent
where the scalar potential $U(x)$ is a Heaviside step function used to create the potential barrier for tunnelling (see Fig.~\ref{fig:1D_tunnelling}\textbf{a}), $i\eta(x) = 4\mathrm{tanh}((|x|/9)^{5})$ is a smooth absorption potential near the boundaries and $\{s_1,s_2,s_3,s_4\}$ are the excitation terms defined by
\begin{equation}\label{eq:transport_source}
s_j(x,y) = \Phi_j(y) e^{ik_x(x-x_s)} e^{-((x-x_s)/w_s)^2}, j=1,2,3,4
\end{equation}
\noindent
where $\Phi(y) = \big[\Phi_1,\Phi_2,\Phi_3,\Phi_4 \big]^T$ is the right-moving DBC mode profile obtained by numerically solving the weak form PDEs (see Eq.~\ref{eq:8}) in one dimension. The source position and width are chosen as $x_s = -5$ and $w_s = 2$. The phases of regions $y>0$ and $y<0$ are the same as those in Fig.~\ref{fig:acoustic}\textbf{b}.

As shown in Fig.~\ref{fig:1D_tunnelling}\textbf{a}, the scalar term $U(x)$ lifts the Dirac point position by $\Delta E = 1$, creating a potential barrier between $E\in[0,1]$. However, this potential barrier does not block the transmission of DBC mode, as shown in the waveguide profile in Fig.~\ref{fig:1D_tunnelling}\textbf{b}. This tunnelling can also be visualized through the transmittance spectrum in Fig.~\ref{fig:1D_tunnelling}\textbf{c}. The transmittance within the barrier-frequency region remains the same as that outside the barrier.

 \begin{figure}
   \centering
   \includegraphics[width=0.7\textwidth]{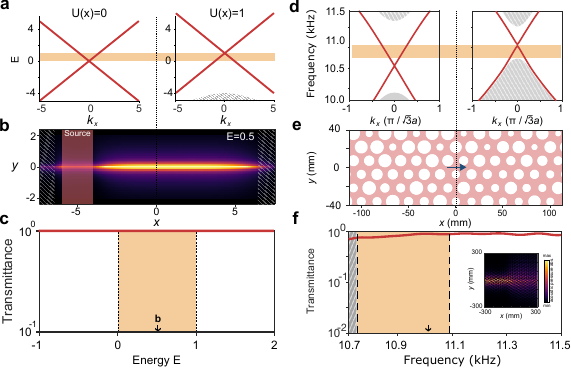}
   \caption{\textbf{The Klein tunnelling of DBC mode.} \textbf{a}, The DBC mode spectrum corresponding to regions with scalar potentials $U(x)=0$ and $U(x)=1$. \textbf{b}, The DBC mode transport profile at $E=0.5$. The red and shaded regions denote source and absorption potential. \textbf{c}, The transmittance spectrum of DBC modes across the potential barrier. \textbf{d}, The spectrum of DBC waveguides with modulation parameters of $R_0=0.2a$ and $R_0=0.17a$. All the other parameters remain the same. \textbf{e}, The DBC waveguide with potential barrier at $x>0$. \textbf{f}, The tunnelling transmittance. Inset is the wave profile obtained at $11$~kHz and indicated by the arrow. The shaded regions denote the frequency regions of potential barrier.}
   \label{fig:1D_tunnelling}
 \end{figure}
 
In the acoustic waveguide structure, we can also create a potential barrier by adjusting the modulation parameter $R_0$ in Eq.~(\ref{eq:Kekule_acoustic}). For example, by introducing an offset of $-0.03a$ to the $R_0$, one can create a potential barrier between $10.5$~kHz and $11.08$~kHz, as shown in Fig.~\ref{fig:1D_tunnelling}\textbf{d}. We further construct an acoustic waveguide in Fig.~\ref{fig:1D_tunnelling}\textbf{e}, analogous to that in Fig.~\ref{fig:1D_tunnelling}\textbf{b}. The modulation parameter $R_0$ for $x>0$ is assumed as $R_0=0.17a$, but remains the same as in the main text for $x<0$; all other design parameters are identical to the main text. In the corresponding transmittance spectrum in Fig.~\ref{fig:1D_tunnelling}\textbf{f}, we observe vanishing reflectance from the potential barrier. This agrees with the experimental results obtained for the spiral waveguide in Fig.~\ref{fig:spiral_waveguide}\textbf{c}.

\section{Freeform waveguiding comparison with acoustic crystal defect waveguide}

In this section, we compare the freeform waveguiding of DBC modes and acoustic crystal (AC) defect modes. The DBC spiral waveguide is created with the same parameters as those in Fig.~\ref{fig:spiral_waveguide}\textbf{a}. The conventional acoustic-crystal (AC) defect consists of a square lattice of solid pillars with lattice constant $p=18.4$ mm and pillar radius $r=7$mm; the defect channel is formed by omitting one line of pillars and enlarging the pillars immediately adjacent to the defect to $1.2r$. The spiral path is identical to that in Fig.~\ref{fig:spiral_waveguide}. As shown in Fig.~\ref{fig:spiral_compar}\textbf{a}, the DBC mode can achieve a better waveguiding performance, regarding both the transmittance amplitude and bandwidths. In contrast, the AC defect waveguide can only realize waveguiding at several individual frequencies. The DBC mode has an overall uniform distribution along the waveguide, which has been shown in simulations (Fig.~\ref{fig:spiral_compar}\textbf{c}) and demonstrated experimentally (Fig.~\ref{fig:spiral_compar}\textbf{d}). In contrast, as shown in Fig.~\ref{fig:spiral_compar}\textbf{b}, the AC defect mode profile is non-uniform along the path, consistent with stronger backscattering in this implementation.

 \begin{figure}
   \centering
   \includegraphics[width=0.95\textwidth]{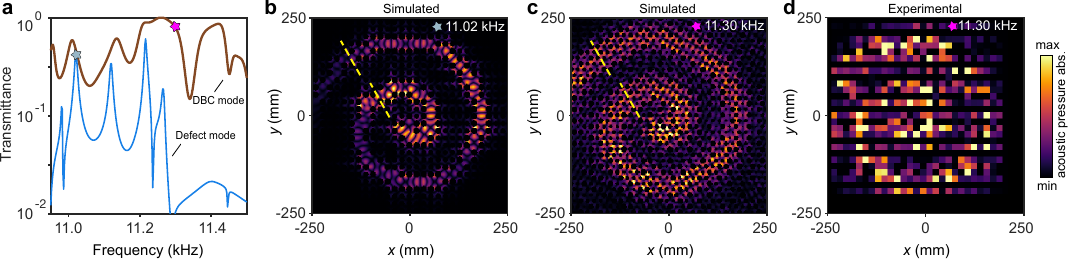}
   \caption{\textbf{The freeform waveguiding of DBC and acoustic crystal defect modes.} \textbf{a}, Transmittance (between the yellow dashed lines in \textbf{b} and \textbf{c}) of DBC and acoustic crystal (AC) defect modes along a spiral path. The parameters for DBC waveguide is same as that in Fig.~\ref{fig:spiral_waveguide} of the main text. The acoustic-crystal defect waveguide is created from a square lattice of solid pillars with lattice constant $p=18.4$ mm and pillar radius $r=7$mm; the defect channel is formed by omitting one line of pillars and enlarging the pillars immediately adjacent to the defect to $1.2r$.  The brown and blue curves denote the transmittance of DBC and AC defect modes. \textbf{b}, The mode profile of guided AC defect mode at $11.02$~kHz. \textbf{c--d}, Simulated and experimental mode profiles of the DBC mode at $11.30$~kHz.}
   \label{fig:spiral_compar}
 \end{figure}

\vskip 0.1in 
\section{Photonic analogue}

In this section, we present FEM simulation results showing that the present scheme for realizing DBC modes in acoustics can be transferred to photonics.

In the photonic design, the acoustic structure's pillars are converted into air holes, while the air regions are converted into GaAs (see Fig.~\ref{fig:photonic}\textbf{a}).  With a lattice constant of $a=28\mu\textrm{m}$ and GaAs refractive index of 3.74, the PhC supports a transverse magnetic (TM) band gap near 3.8 THz, as shown in Fig.~\ref{fig:photonic}\textbf{b}.  For a straight DBC waveguide, analogous to the acoustic design of Fig.~2 in the main text, this yields the dispersion relation shown in Fig.~\ref{fig:photonic}\textbf{b}.  Similar to the acoustic case, we can use the DBC modes to form various configurations such as line cavities, as shown in Fig.~\ref{fig:photonic}\textbf{c}--\textbf{d}.  Photonic realizations at higher frequency regimes can also be achieved by scaling down the dimensions of the structure.

 \begin{figure}
   \centering
   \includegraphics[width=0.9\textwidth]{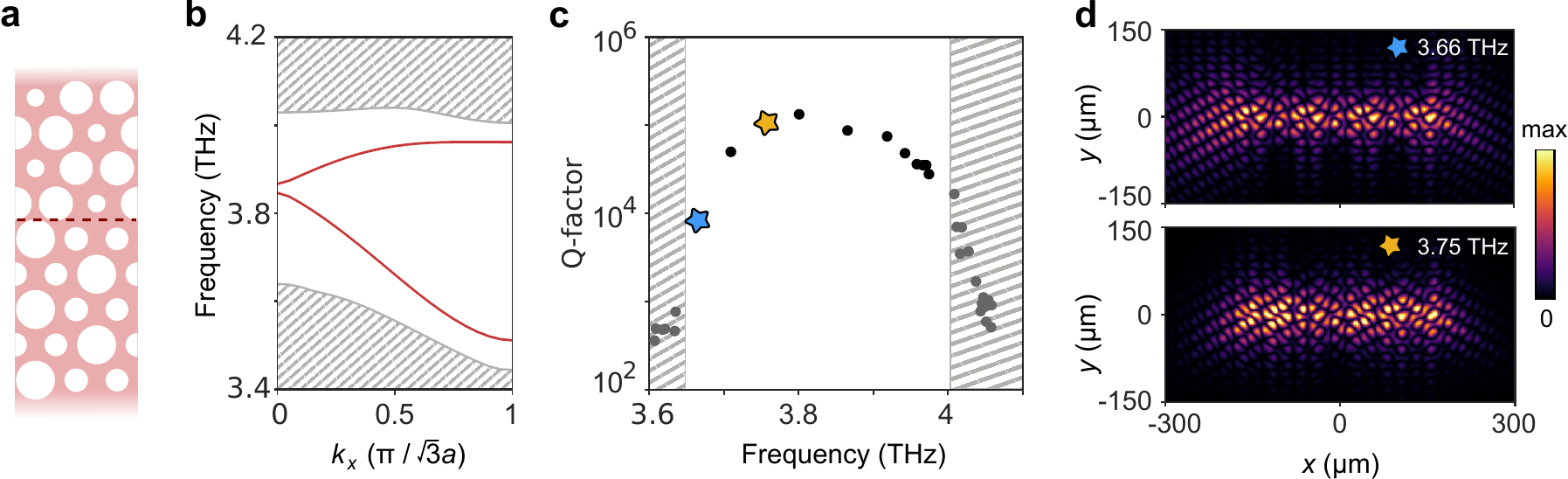}
   \caption{\textbf{Photonic DBC modes.} \textbf{a}, Terahertz-scale photonic structure, similar to the acoustic design but with air holes in a GaAs medium.  The lattice constant is $a$=28 $\mu m$, and the refractive index of GaAs is 3.74.  The domain wall is marked by dashes.  \textbf{b}, Dispersion relations for the transverse magnetic (TM) photonic DBC modes, calculated from FEM simulations.  \textbf{c},  Q-factors for the eigenmode of a DBC line cavity, constructed similarly to Fig.~3 of the main text, with a cavity length 360 $\mu m$.  \textbf{d}, Field amplitude ($|E_z|$) distributions for two of the eigenmodes in \textbf{c}. }
   \label{fig:photonic}
 \end{figure}
 
\clearpage
\end{widetext}


\begin{thebibliography}{1}

\bibitem{Thaller2013}
  B.~Thaller,
  \textit{The Dirac equation}. (Springer Science and Business Media, 2013).

\bibitem{Gottfried1986}
  K.~Gottfried and V.~F.~Weisskopf,
  \textit{Concepts of particle physics}. (Oxford University Press, 1986).

\bibitem{Wehling2014}  
  T.~O.~Wehling, A.~M.~Black-Schaffer and A.~V.~Balatsky,
  Dirac materials,
  Adv.~Phys.~, \textbf{63}, 1 (2014).   

\bibitem{Novoselov2005}
  K.~S.~Novoselov, A.~K.~Geim, S.~V.~Morozov, D.~Jiang, M.~I.~Katsnelson, I.~V.~Grigorieva, S.~V.~Dubonos and A.~A.~Firsov, 
  Two-dimensional gas of massless Dirac fermions in graphene,
  Nature~\textbf{438}, 197 (2005).
  
\bibitem{Neto2009}  
  A.~H.~Castro~Neto, F.~Guinea, N.~M.~R.~Peres, K.~S.~Novoselov and A.~K.~Geim, 
  The electronic properties of graphene,
  Rev.~Mod.~Phys, \textbf{81}, 109 (2009).   

\bibitem{Klein1929}  
  O.~Klein,
  Die reflexion von elektronen an einem potentialsprung nach der relativistischen dynamik von dirac,
  Z.~Physik \textbf{53}, 157 (1929).

\bibitem{Katsnelson2006}  
  M.~I.~Katsnelson, K.~S.~Novoselov and A.~K.~Geim,
  Chiral tunnelling and the Klein paradox in graphene,
  Nat.~Phys.~\textbf{2}, 620 (2006).   

\bibitem{Ni2018} 
  X.~Ni, D.~Purtseladze, D.~Smirnova, A.~Slobozhanyuk, A.~Alù and A.~B.~Khanikaev,
  Spin- and valley-polarized one-way Klein tunneling in photonic topological insulators,
  Sci.~Adv.~\textbf{4}, eaap8802 (2018).    

\bibitem{Jiang2020}
  X.~Jiang, C.~Shi, Zhenglu Li, S.~Wang, Y.~Wang, S.~Yang, S.~G.~Louie and X.~Zhang,
  Direct observation of Klein tunneling in phononic crystals,
  Science 370, 1447 (2020).

\bibitem{Yu2024}
  L.~Yu, H.~Xue, R.~Guo, E.~A.~Chan, Y.~Y.~Terh, C.~Soci, B.~Zhang and Y.~D.~Chong, 
  Dirac mass induced by optical gain and loss,
  Nature~\textbf{632}, 63 (2024).

\bibitem{Jackiw1976} 
  R.~Jackiw and C.~Rebbi,
  Solitons with fermion number $\frac{1}{2}$,
  Phys.~Rev.~D~\textbf{13}, 3398 (1976). 
  
\bibitem{Su1979}
  W.~P.~Su, J.~R.~Schrieffer, and A.~J.~Heeger,
  Solitons in Polyacetylene,
  Phys.~Rev.~Lett.~\textbf{42}, 1698 (1979). 

\bibitem{hasan2010colloquium}
  M.~Z.~Hasan, C.~L.~Kane,
  Colloquium: topological insulators,
  Rev.~Mod.~Phys.~\textbf{82}, 3045 (2010).
  
\bibitem{Qi2011Review}
  X.~L.~Qi, S.-C.~Zhang,
  Topological insulators and superconductors,
  Rev.~Mod.~Phys.~\textbf{83}, 1057 (2011).

%
%
%
%

\bibitem{OzawaReview2019} 
  T.~Ozawa, H.~Price, A.~Amo, N.~Goldman, M.~Hafezi, L.~Lu, M.~Rechtsman, D.~Schuster, J.~Simon, O.~Zilberberg and I.~Carusotto,
  Topological photonics,
  Rev.~Mod.~Phys.~\textbf{91}, 015006 (2019).  

\bibitem{MaReview2019} 
  G.~Ma, M.~Xiao, and C.~T.~Chan,
  Topological phases in acoustic and mechanical systems,
  Nat.~Rev.~Phys.~\textbf{1}, 281 (2019).

\bibitem{ChenReview2026} 
  Z.~Chen, T.~Zhang, X.~Wang, J.~Li, Z.~Lin, F.~Gao, L.~Wang, Y.~Liu, Q.~Wang, X.~Zhang, G.~Ma, X.~Chen, M.~Lu, Y.~Chen and J.~Jiang,
  Topological phononics,
  arXiv:2605.20900v1, (2026).
  
%
%
%
%
%
%
%
%
%
%
%
%
%
%

\bibitem{Jackiw1981}  
  R.~Jackiw and P.~Rossi,
  Zero modes of the vortex-fermion system,
  Nucl.~Phys.~B~\textbf{190}, 681 (1981).  

\bibitem{Weinberg1981}
  E.~J.~Weinberg,
  Index calculations for the fermion-vortex system,
  Phys.~Rev.~D 24, 2669 (1981).

\bibitem{Volovik1999} 
  G.~E.~Volovik,
  Fermion zero modes on vortices in chiral superconductors,
  Pisma~Zh.~Eksp.~Teor.~Fiz.~\textbf{70}, 601 (1999). 

\bibitem{Read2000} 
  N.~Read and D.~Green,
  Paired states of fermions in two dimensions with breaking of parity and time-reversal symmetries and the fractional quantum Hall effect,
  Phys.~Rev.~B~\textbf{61}, 10267 (2000). 

\bibitem{Fu2008} 
  L.~Fu and C.~L.~Kane,
  Superconducting Proximity Effect and Majorana Fermions at the Surface of a Topological Insulator,
  Phys.~Rev.~Lett.~\textbf{100}, 096407 (2008).   
  
\bibitem{Kekule}
  A.~Kekul{\'e},
  Untersuchungen über aromatische Verbindungen Ueber die Constitution der aromatischen Verbindungen. I. Ueber die Constitution der aromatischen Verbindungen.,
  Ann.~Chem.~Pharm.~\textbf{137}, 129 (1866).   
  
\bibitem{Hou2007} 
  C.~Hou, C.~Chamon, and C.~Mudry,
  Electron Fractionalization in Two-Dimensional Graphenelike Structures,
  Phys.~Rev.~Lett.~\textbf{98}, 186809 (2007).   

\bibitem{Gutierrez2016}
  C.~Guti\'errez, C.-J.~Kim, L.~Brown, T.~Schiros, D.~Nordlund, E.~B.~Lochocki, K.~M.~Shen, J.~Park, and Abhay N. Pasupathy,
  Imaging chiral symmetry breaking from Kekul\'e bond order in graphene,
  Nat.~Phys.~\textbf{12}, 950 (2016).

\bibitem{Gao2019} 
  P.~Gao, D.~Torrent, F.~Cervera, P.~San-Jose, J.~Dehesa, and J.~Christensen,
  Majorana-like Zero Modes in Kekul{\'e} Distorted Sonic Lattices,
  Phys.~Rev.~Lett.~\textbf{123}, 196601 (2019).

\bibitem{Noh2020}
  J.~Noh, T.~Schuster, T.~Iadecola, S.~Huang, M.~Wang, K.~P.~Chen, C.~Chamon, and M.~C.~Rechtsman,
  Braiding photonic topological zero modes,
  Nat.~Phys.~\textbf{16}, 989 (2020).
  
\bibitem{Gao2020} 
  X.~Gao, L.~Yang, L.~Hao, L.~Zhang, J.~Li, F.~Bo, Z.~Wang and L.~Lu,
  Dirac-vortex topological cavities,
  Nat.~Nanotechn.~\textbf{15}, 1012 (2020).   

\bibitem{Menssen2020_braiding}
  A.~J.~Menssen, J.~Guan, D.~Felce, M.~J.~Booth and I.~A.~Walmsley,
  Photonic Topological Mode Bound to a Vortex,
  Phys.~Rev.~Lett.~\textbf{125}, 117401 (2020). 

\bibitem{Yang2022} 
  L.~Yang , G.~Li , X.~Gao and L.~Lu,
  Topological-cavity surface-emitting laser,
  Nat.~Photon.~\textbf{16}, 279 (2022). 
  
\bibitem{Han2023} 
  S.~Han, Y.~Chua, Y.~Zeng, B.~Zhu, C.~Wang, B.~Qiang, Y.~Jin, Q.~Wang, L.~Li, A.~G.~Davies, E.~H.~Linfield, Y.~Chong, B.~Zhang and Q.~J.~Wang,
  Photonic Majorana quantum cascade laser with polarization-winding emission,
  Nat.~Commun.~\textbf{14}, 707 (2023).  

\bibitem{martin1966complex}
  D.~Martin, and L.~Ahlfors,
  Complex analysis,
  New York: McGraw-Hill, (1966).

\bibitem{Brown2009}
  J.~W.~Brown and R.~V.~Churchill
  \textit{Complex variables and applications},~8th ed. (McGraw-Hill 2009).

\bibitem{Chen2023}  
  K.~Chen, F.~Komissarenko, D.~Smirnova, A.~Vakulenko, S.~Kiriushechkina, I.~Volkovskaya, S.~Guddala, V.~Menon, A.~Alù and A.~B.~Khanikaev,
  Photonic Dirac cavities with spatially varying mass term,
  Sci.~Adv.~\textbf{9}, eabq4243 (2023).    

\bibitem{Kiriushechkina2023} 
  S.~Kiriushechkina, A.~Vakulenko, D.~Smirnova, S.~Guddala, Y.~Kawaguchi,F.~Komissarenko, M.~Allen, J.~Allen, A.~B.~Khanikaev,
  Spin-dependent properties of optical modes guided by adiabatic trapping potentials in photonic Dirac metasurfaces,
  Nat.~Nanotechn.~\textbf{18}, 875 (2023).     
  
\bibitem{Vakulenko2023}  
  A.~Vakulenko, S.~Kiriushechkina, D.~Smirnova, S.~Guddala, F.~Komissarenko, A.~Alù, M.~Allen, J.~Allen and A.~B.~Khanikaev,
  Adiabatic topological photonic interfaces,
  Nat.~Commun.~\textbf{14}, 4629 (2023).  

\bibitem{Choi2025} 
  Y.~S.~Choi, K.~Y.~Lee, S.~An, M.~Jang, Y.~Kim, S.~Yoon, S.~H.~Shin and J.~W.~Yoon,
  Topological beaming of light: proof-of-concept experiment,
  Light.~Sci.~Appl.~\textbf{14}, 121 (2025).

\bibitem{Mitchell2018}
  N.~P.~Mitchell, L.~M.~Nash, D.~Hexner, A.~M.~Turner, and W.~T.~M.~Irvine,
  Amorphous topological insulators constructed from random point sets,
  Nat.~Phys.~\textbf{14}, 380 (2018).

\bibitem{Wang2021}
  Q.~Wang, Y.~Ge, H.~Sun, H.~Xue, D.~Jia, Y.~Guan, S.~Yuan, B.~Zhang, and Y.~D.~Chong,
  Vortex states in an acoustic Weyl crystal with a topological lattice defect,
  Nat.~Comm.~\textbf{12}, 3654 (2021). 

\bibitem{Banerjee2025}
  R.~Banerjee, A.~Kumar, T.~C.~Tan, M.~Gupta, R.~Jia, P.~Szriftgiser, G.~Ducournau, Y.~Chong, and R.~Singh,
  On-chip amorphous terahertz topological photonic interconnects,
  Sci.~Adv.~\textbf{11}, eadu2526 (2025).

\bibitem{Fleury2023}
  Z.~Zhang, P.~Delplace, and R.~Fleury,
  Anomalous topological waves in strongly amorphous scattering networks,
  Sci.~Adv.~\textbf{9}, eadg3186(2023).
  
      




  



  

  
    
  


\bibitem{Bernevig2006}
  B.~A.~Bernevig, T.~L.~Hughes, and S.-C.~Zhang,
  Quantum Spin Hall Effect and Topological Phase Transition in HgTe Quantum Wells,
  Science 314, 1757 (2006).
  
\bibitem{Torrent2012} 
  D.~Torrent and J.~S´anchez-Dehesa,
  Acoustic Analogue of Graphene: Observation of Dirac Cones in Acoustic Surface Waves,
  Phys.~Rev.~Lett.~\textbf{108}, 174301 (2012). 

\bibitem{He2016_QSH}  
  C.~He, X.~Ni, H.~Ge, X.~Sun, Y.~Chen, M.~Lu, X.~Liu, and Y.~Chen,
  Acoustic topological insulator and robust one-way sound transport,
  Nat.~Phys.~\textbf{12}, 1124 (2016).   
  
\bibitem{Lu2017_valley}  
  J.~Lu, C.~Qiu, L.~Ye, X.~Fan, M.~Ke, F.~Zhang and Z.~Liu,
  Observation of topological valley transport of sound in sonic crystals,
  Nat.~Phys.~\textbf{13}, 369 (2017).   
  
\bibitem{Xu2023} 
  S.~Xu, Y.~Wang and R.~Agarwal,
  Absence of topological protection of the interface states in $\mathbb{Z}_2$ photonic crystals,
  Phys.~Rev.~Lett.~\textbf{131}, 053802 (2023). 

\bibitem{Vahala2003}
  K.~J.~Vahala, Optical microcavities, Nature \textbf{424}, 839 (2003).
 
\bibitem{Song2017PRL}
  Z.~Song, Z.~Fang and C.~Fang,
  $(d\ensuremath{-}2)$-Dimensional Edge States of Rotation Symmetry Protected Topological States,
  Phys.~Rev.~Lett.~\textbf{119}, 246402 (2017). 
  
\bibitem{Benalcazar2017PRB}
  W.~A.~Benalcazar, B.~A.~Bernevig, and T.~L.~Hughes,
  Electric multipole moments, topological multipole moment pumping, and chiral hinge states in crystalline insulators,
  Phys.~Rev.~B~\textbf{96}, 245115 (2017). 

\bibitem{Peterson2018}
  C.~W.~Peterson, W.~A.~Benalcazar, T~L.~Hughes and G.~Bahl,
  A quantized microwave quadrupole insulator with topologically protected corner states,
  Nature~\textbf{555}, 346 (2018). 

\bibitem{Noh2018}
  J.~Noh, W.~A.~Benalcazar, S.~Huang, M.~J.~Collins, K.~P.~Chen, T.~L.~Hughes and M.~C.~Rechtsman,
  Topological protection of photonic mid-gap defect modes,
  Nat.~Photon.~\textbf{12}, 408 (2018).
     





\bibitem{Wu2015} 
  L.~H.~Wu and X.~Hu,
  Scheme for Achieving a Topological Photonic Crystal by Using Dielectric Material,
  Phys.~Rev.~Lett.~\textbf{114}, 223901 (2015).

\bibitem{Joannopoulos2008}
  J.~D.~Joannopoulos, S.~G.~Johnson, J.~N.~Winn and R.~D.~Meade,
  \textit{Photonic crystals: molding the flow of light},~2nd ed. (Princeton University Press, 2008).
  
\end{thebibliography}
\end{document}